\documentclass[preprint,prd,showpacs,superscriptaddress,amsmath,nofootinbib]{revtex4}

\usepackage{cancel}
\usepackage{graphicx}
\usepackage{color}
\usepackage{amssymb}
\usepackage{feynarts}

\newlength{\nseparation}
\newenvironment{nfigure}[1]
        {\begin{figure}[#1]\hrule\vspace{\nseparation}\par}
        {\vspace{\nseparation}\par \hrule \end{figure}}

\newcommand{\bea}{\begin{eqnarray}}
\newcommand{\eea}{\end{eqnarray}}

\newcommand{\eq}[1]{Eq.~(\ref{#1})}

\newcommand{\Msusy}{M_{\rm{SUSY}}}

\newcommand{\ps}{p\hspace{-0.44em}/\hspace{0.06em}}

\begin{document}

\title{Two-loop supersymmetric QCD corrections to Higgs-quark-quark couplings in the generic MSSM \bigskip}

\author{Andreas Crivellin} \email{crivellin@itp.unibe.ch}

\author{Christoph Greub}\email{greub@itp.unibe.ch}
\affiliation{Albert Einstein Center for Fundamental Physics, Institute
  for Theoretical Physics,\\ University of Bern, CH-3012 Bern,
  Switzerland.\bigskip}

\date{\today\bigskip\bigskip}

\begin{abstract}
In this article we compute the two-loop supersymmetric QCD corrections to
Higgs-quark-quark couplings in the generic MSSM generated by diagrams
involving squarks and gluinos. We give analytic results for the two-loop
contributions in the limit of vanishing external momenta for general
SUSY masses valid in the MSSM with general flavor structure. 

Working in the decoupling limit ($M_{\rm{SUSY}} \gg v$) we resum all chirally
enhanced corrections (related to Higgs-quark-quark couplings) up to order $\alpha_s^{(n+1)}
\tan^{n}\beta$. This resummation allows for a more precise determination
of the Yukawa coupling and CKM elements of the MSSM superpotential necessary for the study of Yukawa coupling unification.

The knowledge of the Yukawa couplings of the MSSM superpotential in addition allows us to derive the effective Higgs-quark-quark couplings entering FCNC processes. These effective vertices can in addition be used for the
calculation of Higgs decays into quarks as long as 
$M_{\rm SUSY} > M_{\rm  Higgs}$ holds. Furthermore, our calculation is also necessary for
consistently including the chirally 
enhanced self-energy contributions into the calculation of FCNC processes in the MSSM beyond leading order.

At two-loop order, we find an enhancement of the SUSY threshold corrections, induced by
the quark self-energies, of approximately $9\%$ for
$\mu=M_{\rm{SUSY}}$ compared to the one-loop result. At the same time,
the matching scale dependence of the effective Higgs-quark-quark couplings is significantly reduced.
\end{abstract}


\maketitle

\section{Introduction}
\label{sec:intro}

In the MSSM diagrams with sfermions and gauginos as virtual particles generate important loop corrections to Higgs-quark-quark couplings. After the spontaneous breaking of $SU(2)_L\otimes U(1)_Y$ at the electroweak scale, the Higgs fields acquire their vacuum expectation values (VEVx),and the genuine vertex corrections to Higgs-quark-quark couplings also generate chirality changing quark self-energies (or self-masses). Thus, there is a one to one correspondence between loop corrections to three-point Higgs-quark-quark functions and quark self-energies: The correction to a Higgs-quark-quark coupling is given by the corresponding chirality-changing self-energy divided by the VEV of the involved Higgs field. 
\medskip

This means that we can simplify the calculation of three-point functions by reducing the problem to the calculation of two-point functions (self-energies). In this way, the self-energy contributions to quark masses can be directly related to effective Higgs-quark-quark couplings which allow for an efficient calculation of the effective Higgs vertices. 
\medskip 

The quark self-energies also modify the relation between the Yukawa
couplings of the MSSM superpotential and the quark masses (extracted from low-energy observables). Especially if $\tan\beta$ (the ratio of the VEVs of the two Higgs fields) is large, these contributions are
generically very large and can be of order one
\cite{Hall:1993gn,Carena:1994bv,Carena:1999py,Bobeth:2001sq}. In an analogous way,
also the relation between the CKM matrix of the superpotential and the
physical one is altered (by chargino-squark diagrams in the MSSM with
MFV \cite{Babu:1999hn,Isidori:2001fv,Dedes:2002er,Buras:2002vd,Hofer:2009xb} and in addition by
squark-gluino diagrams in the general MSSM
\cite{Crivellin:2008mq,Crivellin:2009sd}). Because of these corrections
the physical quark masses and the measured CKM elements no longer
equal the ones that appear in the MSSM superpotential. One says that
these relations are modified by so-called threshold
corrections, i.e.. by the decoupling of heavy particles. Since in Higgs decays Higgs mediated FCNCs (like
$B_{s(d)}$ mixing and $B_{s(d)}\to \mu^+\mu^-$) and in Higgsino
vertices the Yukawa couplings (of the superpotential) and not the
physical quark masses enter, a precise knowledge of these quantities
and thus of the threshold corrections is necessary. Furthermore, in
GUT models with Yukawa coupling unification not the effective Yukawa
coupling of the SM, but rather the Yukawas of the superpotential unify
and the SUSY threshold corrections must be taken into account in order
to judge whether they actually do unify \cite{DiazCruz:2000mn,Antusch:2009gu}. In
conclusion, it is desirable to know the relation between the
parameters of the MSSM superpotential and the physical, i.e.,measurable quantities, very precisely. 
\medskip 

Having the relation between the Yukawa couplings (CKM elements) of the
superpotential and the physical quark masses (physical CKM elements)
at hand, one can calculate the effective Higgs couplings entering FCNC
processes that include the SUSY loop corrections. This is most easily
achieved by matching the MSSM on
the two-Higgs-doublet model of type three (2HDM III). The loop-induced
couplings of quarks to the 
``wrong'' Higgs field, i.e., to the Higgs that is not involved in the
Yukawa term in the superpotential, induce flavor-changing neutral
Higgs couplings after switching to the physical basis in which the
quark mass matrices are diagonal in flavor space. These
effective Higgs couplings can be expressed entirely in terms of the
physical masses and self-energies depending on MSSM parameters. Here a
complication arises because these self-energies must be calculated
using the Yukawa couplings and the CKM elements of the superpotential,
which must have been determined previously in the process of
renormalization by including the loop corrections, i.e.,~by resumming
the threshold corrections. This problem can be solved analytically in
the decoupling limit of the generic MSSM in which the self-energies
are at most linear in the Yukawa couplings \cite{Crivellin:2011jt}.
\medskip

The importance of these threshold corrections and thus of the chirally
enhanced self-energies motivates their calculation at NLO in
$\alpha_s$. In the MSSM with MFV these corrections have been
calculated in Refs.~\cite{Guasch:2003cv,Noth:2010jy},
\cite{Bauer:2008bj} and \cite{Bednyakov:2002sf,Bednyakov:2009wt}. Here
we want to extend this analysis to the MSSM with generic sources of
flavor violation and resum all chirally enhanced effects using the
results of
Refs.~\cite{Crivellin:2008mq,Hofer:2009xb,Crivellin:2010er,Crivellin:2011jt}. In
addition, working in the approximation of vanishing external momenta,
we are able to give relatively simple analytic expressions for the
self-energies, and therefore also the resummation of all chirally
enhanced corrections can be (and is) performed analytically.
\medskip

After discussing the quark self-energies (and their connection to
Higgs-quark-quark couplings in the decoupling limit of the MSSM) in
the next section, we derive the relations between the MSSM Yukawa
couplings and the quark masses at LO in Sec.~\ref{renormalization}. As the
main result of this article we calculate the SQCD contribution to the
chirality-changing self-energy at the two-loop level in
Sec.~\ref{2-loop-SQCD}. 
In Sec.~\ref{RenReg} we discuss the topics of
  Sec.~\ref{renormalization} at NLO.
In Sec.~\ref{EFT} we derive the effective
Higgs-quark-quark couplings and conclude in Sec.~\ref{conclusions}. 
Various appendices summarize the relevant one-loop results.
\medskip 

\section{Quark self-energies, effective Lagrangian and the decoupling limit}

As described in the Introduction, there is a one to one correspondence
between chirality changing self-energies and Higgs-quark-quark
couplings: In the decoupling limit of the MSSM ($M_{\rm{SUSY}}>v$ and $M_{\rm{SUSY}}>p$,
where $p$ is the external momentum) chirality changing self-energies
are proportional to one power of a VEV only, and the corrections to the Higgs-quark-quark couplings can be obtained by dividing the corresponding self-energy by the VEV of the Higgs field involved. Thus, as long as the momentum flowing through the Higgs is small compared to the SUSY masses and the SUSY masses are heavier than the electroweak VEV, the decoupling limit is a valid approximation. In this approximation the calculation of the Higgs-quark-quark three-point function can be reduced to the calculation of quark self-energies. For this reason we will consider the quark self-energies in this section in some detail and discuss the decoupling limit. The analysis is valid independent of the loop order (concerning $\alpha_s$ corrections) at which the self-energies are calculated.
\medskip

In general, it is possible to decompose any quark (or any fermion) self-energy into chirality-flipping and chirality-conserving parts in the following way:
\begin{equation}
\Sigma_{fi}^q(p) = \left( {\Sigma_{fi}^{q\,LR}(p^2) + \ps\Sigma
  _{fi}^{q\,RR}(p^2) } \right)P_R + \left( {\Sigma_{fi}^{q\,RL}(p^2) +
  \ps\Sigma_{fi}^{q\,LL}(p^2) } \right)P_L\,.
\label{self-energy-decomposition}
\end{equation}
Note that the chirality-flipping parts $\Sigma_{fi}^{q\,RL,LR}$ have
dimension mass, while the chirality conserving parts $\Sigma_{fi}^{q\,LL,RR}$ are dimensionless. 
\medskip

In the following we will be interested in the contributions to \eq{self-energy-decomposition}
that involve heavy SUSY particles. The reason for this is that only
these contributions lead to the threshold corrections entering the
relation between the quark masses and the Yukawa couplings of the MSSM
superpotential. It is convenient to work in an effective field theory
in which the part of the effective Lagrangian containing mass terms and kinetic terms for the quarks is given by
\begin{equation}
\renewcommand{\arraystretch}{1.8}
\begin{array}{l}
{\cal{L}}^{\rm eff}_{\bar q q}= -\left(v_q Y^{q_i\star}_{\rm tree}\delta_{fi}+C^{q\,RL}_{fi}\right) O^{q\,RL}_{fi} - \left(v_q Y^{q_i}_{\rm tree}\delta_{fi}+C^{q\,LR}_{fi}\right) O^{q\,LR}_{fi}	\\
\phantom{{\cal{L}}^{\rm eff}=}+\left(\delta_{fi}-C^{q\,RR}_{fi}\right)O^{q\,RR}_{fi}+\left(\delta_{fi}-C^{q\,LL}_{fi}\right)O^{q\,LL}_{fi}\,,\\
\label{Leff2quark}
\end{array}
\end{equation}
with the operators defined as
\begin{equation}
\renewcommand{\arraystretch}{1.4}
\begin{array}{l}
O^{q\,RL}_{fi}=\overline{q_f} P_L q_i\,,\qquad	O^{q\,LL}_{fi}=i\overline {q_f} \cancel{\partial} P_L\,q_i\,,\\
O^{q\,LR}_{fi}=\overline{q_f} P_R q_i\,,\qquad	O^{q\,RR}_{fi}=i\overline {q_f} \cancel{\partial} P_R\,q_i\,.\\
\label{operators}
\end{array}
\end{equation}
Throughout this paper, the Wilson coefficients in the
  effective Lagrangian (\ref{Leff2quark}) (or, equivalently, the
  operators) are renormalized in the $\overline{\rm MS}$ scheme.
  The final results for the Wilson coefficients will be written as an 
  expansion in $g_s$, where $g_s$ is meant to be the $\overline{\rm MS}$
  renormalized strong coupling constant
  of the effective theory, running with six (quark) flavors.
\medskip

In \eq{Leff2quark} the term $-v_q Y^{q_i}_{\rm tree} \, \delta_{fi}$
denotes the part of the Wilson 
coefficient of the operator  $O^{q\,RL}_{fi}$
that is induced at tree level by the Yukawa coupling of the MSSM
superpotential. 
The running of $v_q Y^{q_i}_{\rm tree}$ (and also that of
$C^{q\,RL}_{fi}$) is the same as the one
of the quark mass in the SM (in the $\overline{\rm MS}$ scheme). 
At the matching scale $m_{\rm SUSY}$, $Y^{q_i}_{\rm tree}$ is just the
Yukawa coupling $Y^q$ of the MSSM superpotential\footnote{
The matching calculation for  $Y^{q_i}_{\rm tree}$ is most easily done
by using the $\overline{\rm MS}$ scheme, both on the MSSM side
and on the effective theory side. When working up to order $\alpha_s$, we get 
at the matching scale $m_{\rm SUSY}$:
$Y^{q_i}_{\rm tree}=Y^{q_i}$, where $Y^{q_i}$ denotes the
Higgs-quark-quark coupling of the MSSM in the $\overline{\rm MS}$
scheme. However, it is well known that one should use the $\overline{\rm DR}$-scheme
on the MSSM side, such that supersymmetry is preserved.
This can be achieved by the shift 
$Y^{q_i}=(1+\frac{\alpha_s}{4\pi} \, C_F) Y^{q_i}_{\overline{\rm DR}}$.  
This issue will be considered in more detail in Sec.~\ref{RenReg}.
The matching condition then reads:  
$Y^{q_i}_{\rm tree}=(1+\frac{\alpha_s}{4\pi} \, C_F)
Y^{q_i}_{\overline{\rm DR}}$.}.
Note that $Y^{q_i}_{\rm tree}$ is not the effective Yukawa coupling
of the SM, which instead is 
obtained from the physical quark mass see (\eq{mb_Yb_gen}). 
\medskip

The Wilson coefficients $C^{q\,LR,LR}_{fi}$ and $C^{q\,LL,RR}_{fi}$ in
\eq{Leff2quark} 
contain the effects of heavy particles
only. Self-energy diagrams involving no
heavy SUSY particles, 
i.e. ordinary QCD corrections containing only quarks and gluons, do
not contribute to 
the Wilson coefficients in the matching procedure, because they are the 
same on the full side (the MSSM) and on the effective side (the 2HDM
III or the SM). 
At the matching scale $m_{\rm SUSY}$ we find for the Wilson coefficients of \eq{Leff2quark}, using the results for $\Sigma^{\tilde g\,LL}_{q_f q_i}(0)$ and $\Sigma^{\tilde g\,LR}_{q_f q_i}(0)$ given in \eq{SigmaLR1}:
\begin{equation}
\renewcommand{\arraystretch}{2.4}
\left.\begin{array}{l}
  C_{f i }^{q\,LR}  = \dfrac{\alpha_s}{2\pi}
 W_{fs}^{\tilde q} W_{i + 3,s}^{\tilde q\star} \,
C_F \, m_{\tilde g}
\left( {\dfrac{{x_s^2 \ln \left( {x_s^2 } \right)}}{{1 - x_s^2 }} } \right)\,, \\ 
 C_{fi }^{q\,LL} (0)  = - \dfrac{{ \alpha _s
 }}{{4\pi }}  W_{fs}^{\tilde q} W_{is}^{\tilde q\star} C_F \left( { \ln \left( {x_\mu ^2 } \right) + \dfrac{{3 - 4x_s^2  + x_s^4  + \left( {4x_s^2  - 2x_s^4 } \right)\ln \left( {x_s^2 } \right)}}{{2\left( {1 - x_s^2 } \right)^2 }}} \right)\,, \\ 
  Y^q_{\rm tree}=Y^q\,,
 
 \end{array}\!\right\}\,{\rm at\,LO\, in\,}\alpha_s\,.
 \label{CLO}
\end{equation}

\medskip

Further, in the following we will focus on the nondecoupling pieces of
\eq{self-energy-decomposition}, i.e., those contributions that do not
vanish in the
 limit $M_{\rm SUSY}\to \infty$ (which also includes $\mu\to
 \infty$). In contrast, 
all parts that vanish in this limit are called decoupling. There are
two different kinds of decoupling 
contributions concerning self-energies (or effective Higgs-quark couplings):

\begin{itemize}
	\item The first kind of decoupling effects is related to the
          expansion of the self-energies in powers of 
$p^2/M^2_{\rm{SUSY}}$. This expansion is certainly possible in on-shell configurations because the SUSY particles are known to be much heavier than the external quarks. In this series, higher order contributions are clearly suppressed for all light quarks and even for the top quark, nondecoupling corrections are only of the order $m_t^2/M_{\rm{SUSY}}^2\leq4\%$ with respect to the leading term. Thus, higher orders in $p^2/M_{\rm{SUSY}}^2$ can be safely neglected as long as the external momentum $p^2$ is small, which is the case for all low-energy flavor observables.
	
	\item The second kind of decoupling effect is related to the mixing matrices (and also the physical masses) of the MSSM particles (squarks and charginos/neutralinos) which appear because the mass matrices of the SUSY particles are not diagonal in a weak basis. These mixing matrices and mass eigenvalues can be expanded in powers of $v/M_{\rm{SUSY}}$, and also in this case it turns out that the decoupling limit (i.e., the leading order $v/M_{\rm{SUSY}}$) for realistic values of SUSY masses\footnote{The new results
   of the CMS Collaboration \cite{Collaboration:2011wc} and the
   ATLAS experiment \cite{daCosta:2011qk} require that squark and
   gluino masses are at least of the order of 1 TeV.} is an excellent approximation to the full expressions \cite{Crivellin:2010er}. Beyond the decoupling limit higher dimensional operators involving several Higgs fields would appear. 
\end{itemize}

From dimensional analysis we see that all nondecoupling contributions
are contained in $\Sigma_{fi}^{q\,RR,LL}$ and $\Sigma_{fi}^{q\,LR,RL}$
evaluated at $p^2=0$. Furthermore, the nondecoupling part of
$\Sigma_{fi}^{q\,RR,LL}(p^2=0)$ is independent of a VEV, while 
$\Sigma_{fi}^{q\,LR,RL}(p^2=0)$ is linear in $v$. Thus, in the
following we will work in the limit $\Sigma_{fi}^{q\,RR,LL}(p^2=0)$,
$\Sigma_{fi}^{q\,LR,RL}(p^2=0)$ and only keep the leading term in $v$
that is equivalent to considering operators up to dimension 4
only. This simplification allows us to perform an analytic resummation
of all chirally enhanced effects as developed in Ref.~\cite{Crivellin:2011jt}.
\medskip 

There is a fundamental difference between $\Sigma_{fi}^{q\,LR,RL}$ and
$\Sigma_{fi}^{q\,RR,LL}$ (and thus also between $C_{fi}^{q\,LR,RL}$
and $C_{fi}^{q\,RR,LL}$) even though both pieces do not decouple.
We explain this issue at one-loop order:
$\Sigma_{fi}^{q\,RR,LL}$ enters always proportional to the quark mass
itself into the renormalization of the Yukawa coupling and CKM
elements and thus has the same generic size as an ordinary QCD loop
correction (it is of order $\alpha_s$). Furthermore, as we will see
later, the $\Sigma_{fi}^{q\,RR,LL}$ even do not contribute to
effective Higgs-quark-quark couplings at the one-loop
level~\cite{Gorbahn:2009pp}. On the other hand,
$\Sigma_{fi}^{q\,LR,RL}$ can be ``chirally enhanced'' by a factor of
$\tan\beta$ \cite{Blazek:1995nv} or $A^f_{ij}/(Y^f_{ij}\Msusy)$
\cite{Crivellin:2008mq}, which can compensate for the loop factor. 
Because of this possible enhancement,
$\Sigma_{fi}^{q\,LR,RL}$ generates the most important contribution to the threshold corrections between Yukawa couplings and quark masses.
The resulting Wilson coefficients $C_{fi}^{q\,LR,RL}$ can even be of
order one, i.e. numerically as large as the corresponding physical
quantities ($m_{q_i}$ in the flavor-conserving case or
$V_{fi}\times{\rm max}\left[m_{q_i},m_{q_i}\right]$ in the flavor-
changing one). Furthermore, concerning flavor-changing neutral Higgs
couplings, $\Sigma_{fi}^{q\,LR,RL}$ even constitutes the leading order,
since these couplings are first generated at the one-loop level.
\medskip

Because the gluino contribution to $\Sigma_{fi}^{q\,LR,RL}$ involves
the strong coupling constant, it is the numerically dominant
contribution to the threshold corrections modifying the relations
between the quark masses and the Yukawa coupling. Regarding flavor
changes, in the MSSM with MFV only the chargino contribution enters
the renormalization of the CKM matrix, but once there are sizable
nonminimal sources of flavor violation, again the gluino
contribution becomes dominant. The neutralino contribution is in most
regions of parameter space suppressed (except if the gluino is much
heavier than the other SUSY particles). Thus we consider the gluino
contribution in this article. The calculation of the chargino- and
neutralino-induced contributions to the threshold corrections and the
effective Higgs-quark-quark couplings is work in progress
\cite{Chargino}.

\medskip 

From the arguments given above we see that at any loop order
(concerning $\alpha_s$ corrections) the chirality-flipping quark
self-energy containing at least one gluino and one squark as virtual
particles is always proportional to one\footnote{More precisely, in
  the decoupling limit $\Sigma_{fi}^{q\,LR}$ is linear in
  $\Delta^{d\,LR}$, while beyond the decoupling limit it contains all
  add powers of $\Delta^{d\,LR}$.} off-diagonal element
$\Delta^{q\,LR}_{ij}$ of the squark mass matrix that, in the
super-CKM basis, is given by

\begin{equation}
\renewcommand{\arraystretch}{1.4}
\begin{array}{l}
\Delta^{d\,LR}_{ij}=-v_d A^d_{ij}\;-\;v_u A^{\prime
  d}_{ij}\;-\;v_u\,\mu\, Y^{\tilde d_i}\, \delta_{ij}\,,\\
\Delta^{u\,LR}_{ij}=-v_u A^u_{ij}\;-\;v_d A^{\prime
  u}_{ij}\;-\;v_d\,\mu\, Y^{\tilde u_i}\, \delta_{ij}\,,
  \end{array}
\label{DeltaLR}
\end{equation}
with $\Delta^{q\,RL}_{ij}=\Delta^{q\,LR\star}_{ji}$. Note the presence
of the tilde in the Yukawa couplings $Y^{\tilde q_i}$. This refers to
the fact that a squark-squark-Higgs coupling is involved, while
$Y^{q_i}$ entering the Wilson coefficient $Y^{q_i}_{\rm tree}$ in
\eq{Leff2quark} is a quark-quark-Higgs coupling.
Of course, both of these couplings are \textit{a priori} equal in the
MSSM owing to supersymmetry and could be identified with each other from the beginning if the calculations
of the chirality-flipping quark self-energies would be performed in the
$\overline{\rm DR}$scheme, in which supersymmetry is preserved.
However, we decided to work out in an intermediate step the SQCD two-loop corrections to the
self-energies in the $\overline{\rm MS}$scheme, i.e., 
in dimensional regularization followed by modified
minimal subtraction rather than using dimensional reduction. At this level, 
the two couplings $Y^{q_i}$ and $Y^{\tilde q_i}$ are different and
therefore have to be distinguished in the notation.
We will discuss this in more detail in Sec.~\ref{RenReg}.

\medskip

The elements $\Delta^{q\,LR}_{ij}$ generate
chirality-enhanced effects with respect to the tree-level quark masses if
they involve the large VEV $v_u$ ($\tan\beta$ enhancement for the
down quark) or a trilinear $A^{(\prime)q}$ term
$A^{(\prime)q}_{ij}/(Y^q_{ij}\Msusy)$-enhancement. 
\medskip 

\subsection{Decomposition of quark self-energy contributions}
We diagonalize the full $6\times 6$ squark mass matrices in the
following way\footnote{Note that our mixing matrices $W^{\tilde{q}}$ 
correspond to the Hermitian conjugate of the matrices $\Gamma_Q$ 
defined in Refs.~\cite{Borzumati:1999qt,Besmer:2001cj}.}:
\begin{equation}
W^{\tilde{q}\dagger}\,\mathcal{M}^2_{\tilde{q}} \,W^{\tilde{q}} =
\textrm{diag}(m_{\tilde{q}_1}^2,m_{\tilde{q}_2}^2,
m_{\tilde{q}_3}^2,m_{\tilde{q}_4}^2,m_{\tilde{q}_5}^2,m_{\tilde{q}_6}^2)
\,,
\label{Wdef}
\end{equation}
where $m_{\tilde{q}_s}$ ($s=1,...,6$) denote the physical squark
masses.
\medskip 

In the decoupling limit, i.e., to leading order in $v/\Msusy$, the chirality-flipping elements
$\Delta^{q\,LR}$ can be neglected in the determination 
of the squark mixing matrices $W^{\tilde{q}}$ and the physical squark masses $m_{\tilde{q}_s}^2$. The down (up) squark mass matrices are then block diagonal and diagonalized by the mixing matrices $\Gamma^{ij}_{DL},\Gamma^{ij}_{DR}$ ($\Gamma^{ij}_{UL},\Gamma^{ij}_{UR}$) in the following way:
\begin{eqnarray}
W^{\tilde{q}\dagger}_{\rm dec}\,\mathcal{M}^2_{\tilde{q}}\,W^{\tilde{q}}_{\rm dec}&=&\textrm{diag}(m_{\tilde{q}_1^L}^2,m_{\tilde{q}_2^L}^2,
m_{\tilde{q}_3^L}^2,m_{\tilde{q}^R_1}^2,m_{\tilde{q}^R_2}^2,m_{\tilde{q}^R_3}^2)\,,\hspace{1cm}
W^{\tilde{q}}_{\rm dec}\,=\,\begin{pmatrix} \Gamma_{QL} & 0 \\ 0 & \Gamma_{QR} \end{pmatrix}\,.\label{eq:Wmat}\end{eqnarray}
The $3\times 3$ matrices $\Gamma_{QL}^{ij}$ and $\Gamma_{QR}^{ij}$ ($Q=U,D$) take into
account the flavor mixing in the left-left and right-right sector of sfermions,
respectively.
It is further convenient to introduce the abbreviations
\bea
\Lambda_{m\,ij}^{q\,LL} \,=\, \Gamma_{QL}^{im}\,\Gamma_{QL}^{jm\star}\,,
&\hskip 1cm & (q=u,d), \nonumber\\ 
\Lambda_{m\,ij}^{q\,RR} \,=\, \Gamma_{QR}^{im}\,\Gamma_{QR}^{jm\star}\,,
& &
\label{eq:Vmat}
\eea
where $i,j,m=1,2,3,$ and the index $m$ is not summed over. 

On the other hand, left-right mixing of squarks is
not described by a mixing matrix, but rather treated perturbatively in
the form of two-point $\tilde{q}^R_i$-$\tilde{q}^L_j$ vertices
governed by the couplings $\Delta^{q\,LR}_{ji}$, i.e., by what is called
mass insertions \cite{Hall:1985dx}.
\medskip 

For the relations between the Yukawa couplings and the quark masses (to be discussed in Sec.~\ref{renormalization}) and for the effective Higgs-quark-quark vertices (see Sec.~\ref{EFT}) it is necessary to decompose $C_{ii}^{d\,LR,RL}$ according to its $Y^{d}$ dependence as
\begin{equation}
C_{ii}^{d\,LR} \;=\;
C_{ii\,\cancel{Y_i}}^{d\,LR} \, + \,
\epsilon_i^{d}\,v_u\,\,Y^{\tilde d_i}\,.
\label{eq:epsilon_b}
\end{equation}
where, as the notation implies, $C_{ii\,\cancel{Y_i}}^{d\,LR}$ is
independent of a Yukawa coupling. Note that we did the decomposition
with respect to the Yukawa coupling $Y^{\tilde d_i}$, as
$C_{fi}^{d\,LR}$ can only involve $Y^{\tilde d_i}$ but
 not $Y^{d_i}$ see \eq{DeltaLR}.
\medskip

For the discussion of the effective Higgs-quark-quark vertices in Sec.~\ref{EFT} we also need a decomposition of $\Sigma_{ji}^{q\,LR}$ and thus of
$C_{ji}^{q\,LR}$ into its holomorphic and nonholomorphic parts\footnote{With (non-)holomorphic we mean that the loop induced Higgs coupling is to the (opposite) same Higgs doublet as involved in the corresponding Yukawa coupling of the MSSM superpotential.}. 
In the decoupling limit (and in the approximation $m_q=0$) all holomorphic self-energies are proportional to $A$~terms.  Thus we denote the holomorphic part of the Wilson coefficient as $C_{ji\,A}^{f\,LR}$, while the
nonholomorphic part (which can be induced by the $\mu$ term or by an $A^\prime$ term) is denoted as $C_{ji}^{\prime q \,LR}$.
This means that we have the relation
\begin{equation}
C_{ji}^{q\,LR} = C_{ji\,A}^{q\,LR} + C_{ji}^{\prime q \,LR}\,.  \label{HoloDeco}
\end{equation}

\section{Relations between quark masses and Yukawa couplings at leading order in $\alpha_s$}
\label{renormalization}

Let us discuss the renormalization\footnote{Throughout this article, renormalization is not only understood as the process of removing divergences, but also as the altering of the relations between different quantities induced by loop contributions.} of quark masses and Yukawa couplings induced by nondecoupling self-energy contributions to the Wilson coefficients $C_{ji}^{q \,LR,RL}$ and $C_{ji}^{q \,LL,RR}$ in the MSSM. For this purpose we focus on the flavor-conserving case, but we will return to the flavor-changing one in Sec.~\ref{EFT}. As it turns out, flavor-changing self-energies only contribute to the relation between quark masses and Yukawa couplings at higher orders in the perturbative diagonalization of the quark mass matrices.
\medskip  

For the renormalization and the inclusion of the threshold corrections
it is very important to distinguish between the Yukawa couplings of
the MSSM superpotential $Y^q$ and the ``effective'' Yukawa couplings of
the SM (or the 2HDM of type III) $Y^q_{\rm{eff}}=m_{q_i }/v_q$. At the
matching scale $M_{\rm{SUSY}}$ the running quark mass $m_{q_i}$ of the
SM is related to the Yukawa coupling of the MSSM in the following way: 
\begin{equation}
v_q Y^{q_i}_{\rm{eff}}=m_{q_i} = \left(v_q Y^{q_i}_{\rm tree}  + C_{ii}^{q\;LR}\right)\times\left(1+ \dfrac{1}{2} \left( {C_{ii}^{q\;LL}  + C_{ii}^{q\;RR} } \right)\right)  \,. \label{mb_Yb_gen}
\end{equation}
The term $\dfrac{1}{2} \, \left( {C_{ii}^{q\;LL}  + C_{ii}^{q\;RR} }
\right)$ originates from rendering the kinetic terms of the effective
theory diagonal, or, equivalently in the full theory from the Lehmann-Symanzik-Zimmermann factor that originates for
the truncation of the external legs. 
\medskip  

As discussed in the last section, only $\Sigma _{ii}^{q\;LR}$ (or equivalently $C_{ii}^{q\;LR}$ in the effective theory) can be chirally enhanced. If we restrict ourselves to this term we recover (in the decoupling limit in which $C_{ii}^{q\;LR}$ is proportional to one power of $Y^{d_i}$ at most) the well-known resummation formula for $\tan\beta$-enhanced corrections, with an additional correction attributable to the $A$~terms \cite{Guasch:2003cv} (and possibly the $A'$ terms). The resummation formula at leading order is given by\footnote{For large flavor-changing elements also a contribution involving two self-energies can be important for the renormalization of the light quark masses \cite{Crivellin:2010gw}. In this case the resummation formula reads for $i=1$: $Y^{d_1} = \dfrac{m_{d_1} - C_{11\,\cancel{Y_1}}^{d\,LR}-\dfrac{C_{13}^{d\,LR}C_{31}^{d\,LR}}{m_{d_3}}}{v_d
  \left( {1 + \tan\beta \epsilon_1^d } \right)}$}
\begin{equation}
Y^{d_i} = \dfrac{m_{d_i} - C_{ii}^{d\,LR\,(1)}}{v_d}= \dfrac{m_{d_i} - C_{ii\,\cancel{Y_i}}^{d\,LR\,(1)}}{v_d
  \left( {1 + \tan\beta \epsilon_i^{d\,(1)} } \right)} \,,
\label{md-Yd}
\end{equation}
with $\epsilon_i^{d\,(1)}$ and $C_{ii\,\cancel{Y_i}}^{d\,LR\,(1)}$ defined through \eq{eq:epsilon_b}. The superscript (1) denotes the fact that a corresponding quantity is calculated at the one-loop order. 
\medskip

\section{Calculation of the Wilson coefficient $C^{q\,LR}_{fi}$ at NLO}
\label{2-loop-SQCD}

In this section we describe the calculation of the two-loop contribution
to $C^{q\;LR}_{fi}$, discuss the issue of renormalization, show the
expected reduction of the matching scale dependence and discuss the
decoupling limit in which only one coupling to a VEV of a Higgs field
is involved. To be specific, we describe in the following the calculation and the
results for the down quark, i.e., $C^{d\;LR}_{fi}$, and mention at the very end how
$C^{u\;LR}_{ij}$ can be obtained.
\medskip

\begin{figure}
\centering
\includegraphics[width=0.53\textwidth]{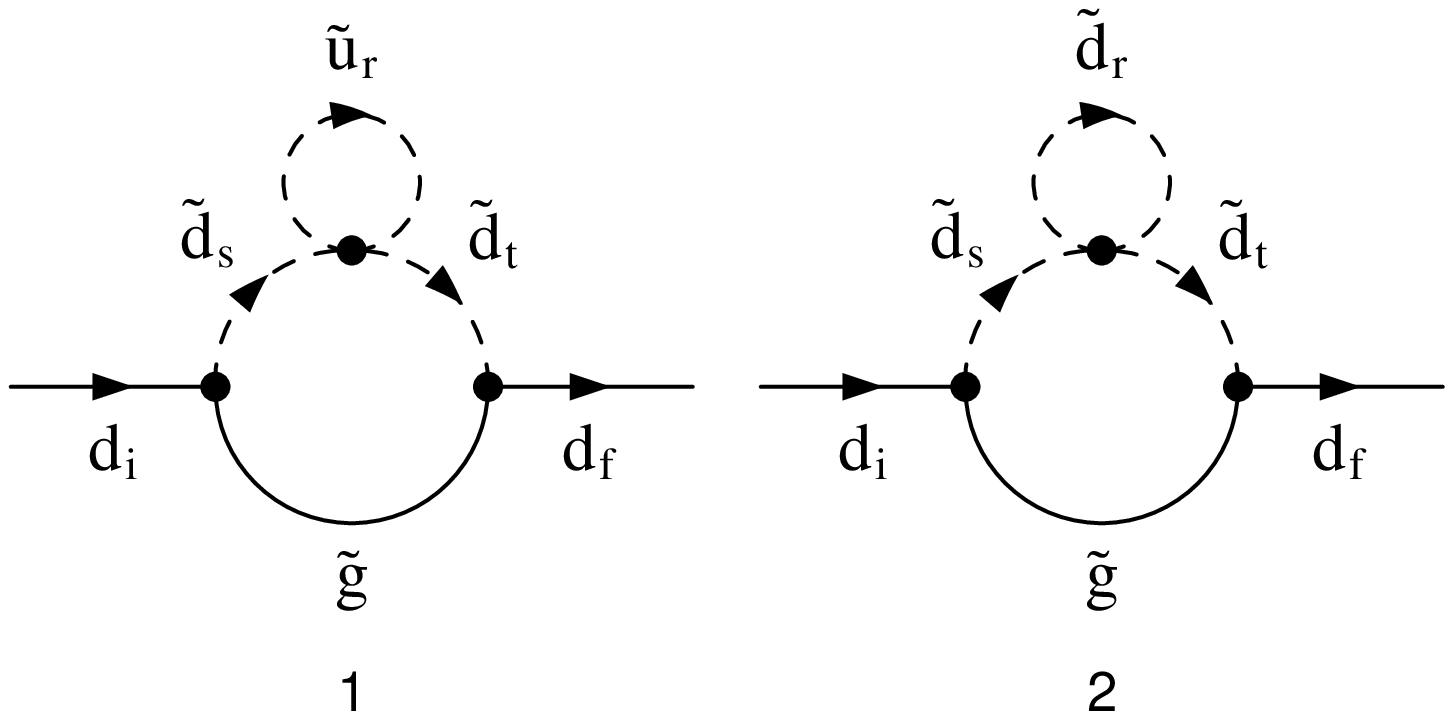}

\includegraphics[width=0.74\textwidth]{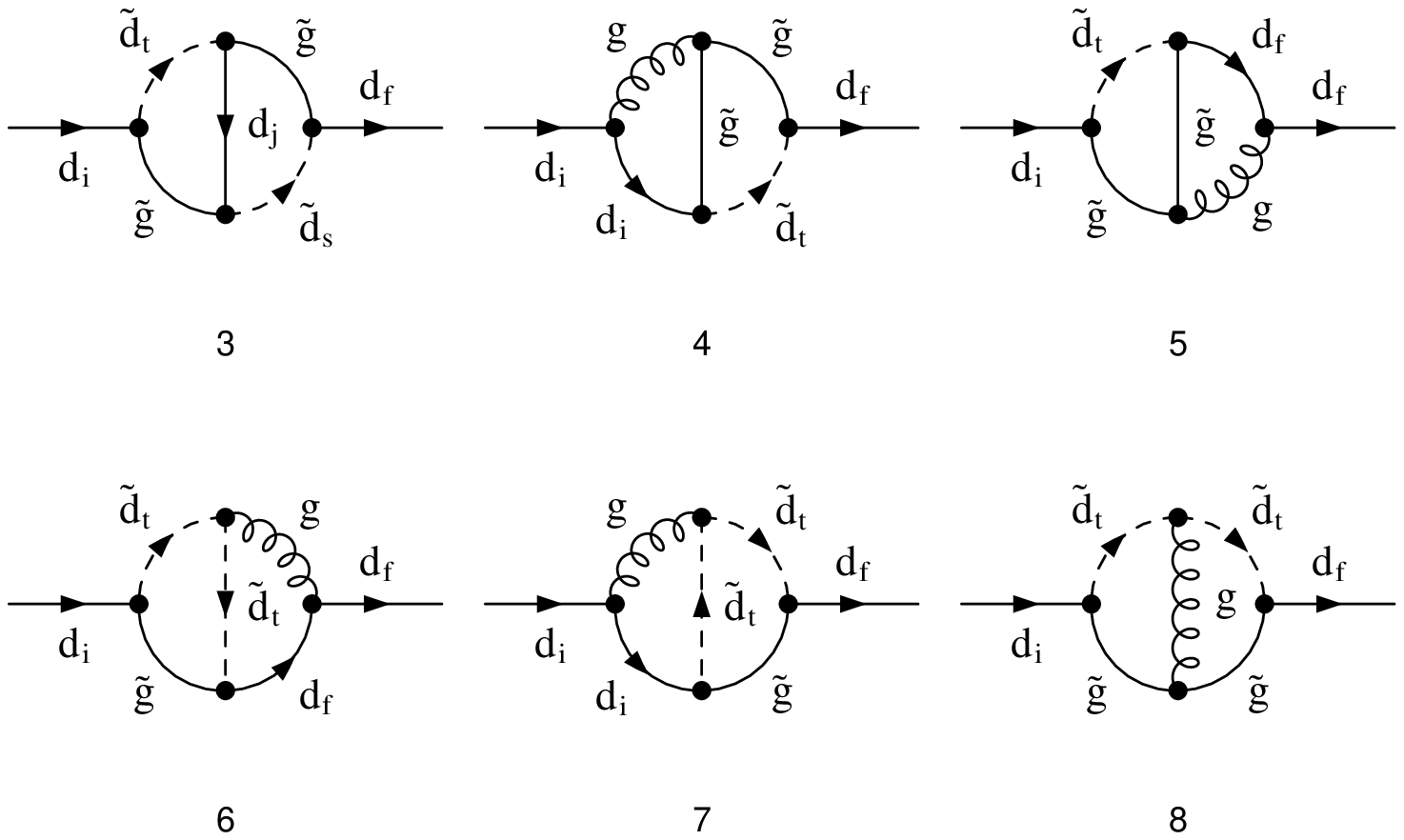}

\includegraphics[width=1\textwidth]{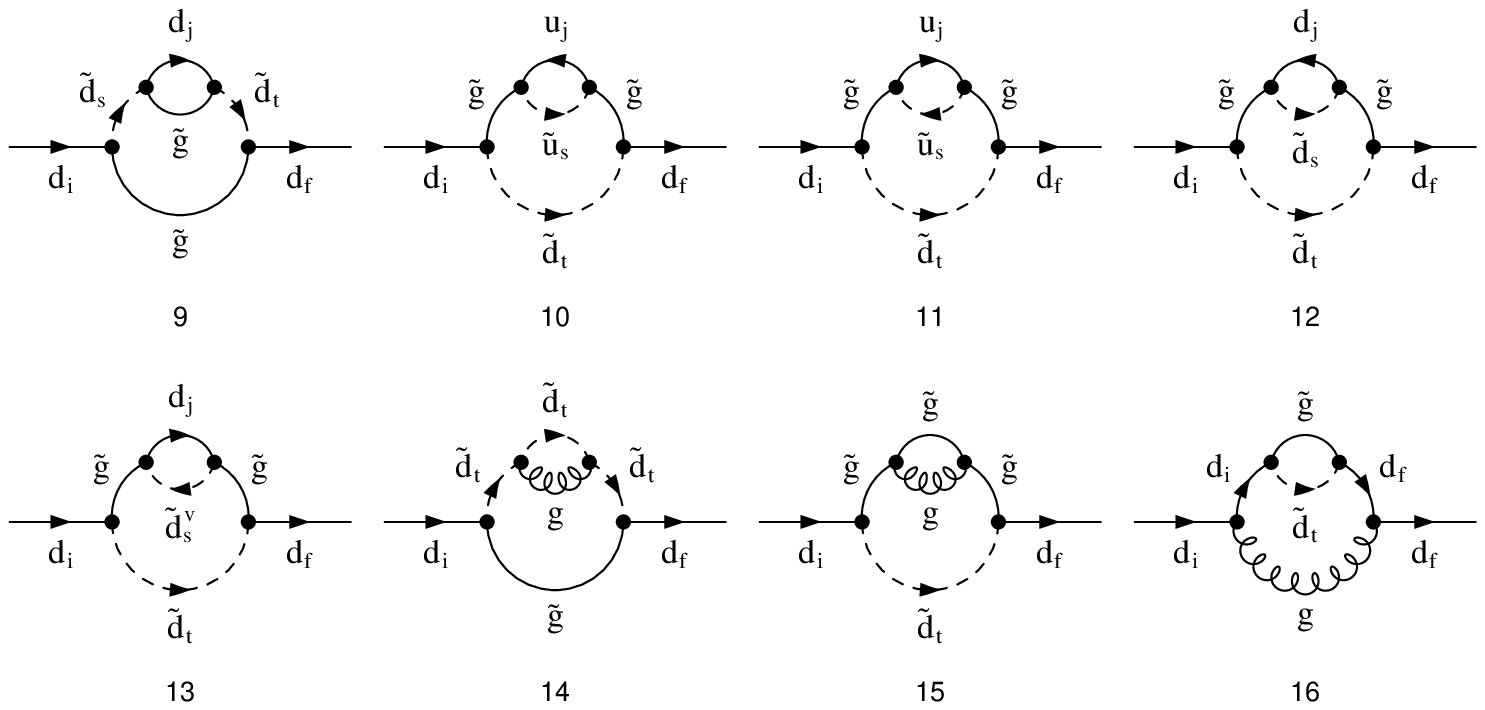}

\caption{Genuine 1-PI two-loop diagrams involving squarks and gluinos necessary for the calculation of $C^{d\,LR\,(2)}_{fi}$.}\hrule\label{2-loop-diagrams}
\end{figure}

In the following we write the Wilson coefficient $C_{fi}^{d\;LR}$ as
\begin{equation}
C_{fi}^{d\;LR} = C_{fi}^{d\,LR\,(1)} + C_{fi}^{d\,LR\,(2)}
+ \ldots \, ,
\label{C1C2}
\end{equation}
where $C_{fi}^{d\,LR\,(1)}$ and $C_{fi}^{d\,LR\,(2)}$ denote the one- and
two-loop contributions, respectively.
We perform the two-loop matching calculation (order $\alpha_s^2$) for the Wilson coefficient
$C^{d\;LR}_{fi}$ in $D=(4-2\varepsilon)$~dimensions, using dimensional
regularization, both for the full theory (MSSM) and for the effective
theory in \eq{Leff2quark}.
The complete list of genuine 1-PI two-loop diagrams contributing in
the full theory is shown in
Fig.~\ref{2-loop-diagrams} (generated with
FeynArts \cite{Hahn:2000kx,Hahn:2001rv}).
 
As the first two diagrams (involving squark tadpoles) give rise to some
subtle points concerning renormalization, we ignore them in this subsection 
and take into account their impact on $C^{d\,LR}_{fi}$ only in the next subsection. 

\subsection{Matching calculation for $C^{d\,LR\,(2)}_{fi}$ ignoring tadpoles}
\label{Sec:ohnetadpole}
In the full theory we first calculate
the 1-PI two-loop diagrams (diagrams 3 - 16 in Fig.~\ref{2-loop-diagrams}) in the approximation $m_q=0$ and $p=0$, but
to all orders in $v/m_{\rm SUSY}$ (using exact diagonalization of the
squark mass matrices). All diagrams except diagram 16 can be
calculated by naively setting $m_q=0$ and $p=0$. Diagram~16, however, leads to two contribution:
the {\it hard} contribution, which amounts to the naive limit of vanishing quark masses and external momenta of the full
two-loop diagram, and the {\it soft} contribution which amounts to the same limit but only for
the heavy one-loop subdiagram \cite{Smirnov:1994tg}. As the soft contribution
is identical to the one-loop gluon correction to 
$-i \, C^{d\,LR\,(1,D)}_{fi} \, O^{d\;LR}_{fi}$ in the effective
theory\footnote{$C^{d\,LR(1,D)}_{fi}$ is the one-loop Wilson
  coefficient in $D=(4-2\varepsilon)$~dimensions, i.e.
$C^{d\,LR\,(1,D)}_{fi}=\Sigma^{\tilde{d} \, LR}_{d_f d_i}(0)$, see
  \eq{SigmaLR1}.}, this contribution drops out in the matching for
$C^{d\,LR\,(2)}_{fi}$. As this soft contribution is the only one that
is infrared singular, this means in particular that
$C^{d\,LR\,(2)}_{fi}$ is free of infrared problems, as it should be.
\medskip

We then add the counterterm contributions in the full theory which are
induced by the renormalization of the parameters $m_{\tilde q_s}^2$,
$m_{\tilde{g}}$ and $\alpha_s$ in the corresponding one-loop result
(where at this level of the calculation these three parameters are renormalized in the 
$\overline{\rm MS}$ scheme).
The explicit expressions are listed in Sec.~\ref{sec:counter-term}. 
In one of these counterterm contributions the squark-mass counterterm 
$\delta m^2_{\tilde{q}_s}$ enters. Of course, when ignoring the tadpole diagrams in this section, the tadpole
contribution to $\delta m^2_{\tilde{q}_s}$ also has to be ignored.

Besides the renormalization of the parameters in the full theory, we
also have to attach one-loop wave function renormalization constants for the
external quark legs to the corresponding one-loop result. These wave
function renormalization constants have two contributions: One from a self-energy
with a gluon-quark loop and another one from a gluino-squark loop.
The first one is also present in the effective theory and consequently drops out in the determination
of $C^{d\,LR\,(2)}_{fi}$, while the second one contributes. Since we perform the renormalization in the $\overline{\rm MS}$ scheme, only the divergent pieces of $\Sigma^{\tilde g\,LL,RR}_{d_f d_i}$ enter $C^{d\,LR\,(2)}_{fi}$ while the finite part gives rise to $C^{d\,LL,RR}_{fi}$.
\medskip

We now turn to the effective theory. Here, we have to work out one-loop QCD
corrections to $-i \, C^{d\;LR(1,d)}_{fi} \, O^{d\;LR}_{fi}$, i.e., 
the 1-PI diagram, attach the wave function renormalization constants and 
take into account the effect of the ($\overline{\rm MS}$) renormalization constant $\delta Z_O$ of the operator
$O^{d\,LR}_{fi}$. While the first two get canceled against
contributions in the full theory (as already mentioned above), the
effect of the renormaliztion constant of the operator enters the
matching condition for  $C^{d\,LR(2)}_{fi}$.
\medskip

Putting things together, we get the following (schematic) matching
equation:
\begin{eqnarray}
-i \, \delta Z_O \, C^{d\;LR(1,d)}_{fi} - i  \, C^{d\;LR(2)}_{fi} = &&
D_3 +\ldots+ D_{15} + D_{16}^{\rm hard} - i \left[ {\rm CT}_{m_{\tilde g}} +
  {\rm CT}_{m_{\tilde{q}_s}} + {\rm CT}_{\alpha_s} \right] \nonumber \\ 
&&- i  \frac{1}{2}
\left[\delta Z^{\rm heavy}_{2,f}
  +\delta Z^{\rm heavy}_{2,i}\right] \, C^{d\;LR(1,d)}_{fi} \, .
\label{matchingschematic}
\end{eqnarray}
Here ${\rm CT}_{m_{\tilde g}}$, ${\rm CT}_{m_{\tilde{q}_s}}$ and ${\rm CT}_{\alpha_s}$ stand for the contributions induced by the insertions of the corresponding counterterms into the one-loop diagram and $D_i$ represents the contribution stemming from diagram $i$ of Fig~\ref{2-loop-diagrams}. As already mentioned, we did our two-loop calculation in dimensional
regularization. So far the parameters $m_{\tilde g}$,
$m_{\tilde{q}_s}$ and $\alpha_s$ appearing in the full theory were
renormalized according to the $\overline{\rm MS}$~scheme.
Also the various $Z-$factors appearing in \eq{matchingschematic} are
renormalized in the $\overline{\rm MS}$ scheme. The result for
$C^{d\;LR(2)}_{fi}$ we get at this level corresponds to the sum of the
first five terms on the right-hand side of \eq{Cdecomp}. When giving
the explicit expressions for these terms, we freely made use of the
unitarity properties of the $W^{\tilde{q}}$ mixing matrices.
\medskip
  
We should be more precise concerning $g_s$ (or $\alpha_s$). In our calculation of the
full theory side $g_s$ stands for $g_{s,Y}$, i.e. for the strong coupling constant of the
Yukawa type of the full MSSM renormalized in the $\overline{{\rm MS}}$
scheme. As we want to express the final result for the Wilson
coefficient $C^{d\,LR\,(2)}_{fi}$
in terms of $g_{s,\overline{{\rm MS}}}^{(6)}$, i.e. by
the strong coupling constant of the SM in the 
$\overline{{\rm MS}}$~scheme running with six flavors, we 
make use of the relation \cite{Martin:1993yx,Mihaila:2009bn}
\begin{equation}
\alpha_{s,Y}(\mu) = \left\{ 1 + \dfrac{\alpha_s}{4\pi} \dfrac{1}{3}
\left[(n_f +6) \ln(x_\mu^2) - \sum \limits_{s=1}^6
  \left(\ln(x_s)+\ln(y_s)\right) + 4 \, C_A - 3\, C_F \right] \right\}
\alpha_{s,\overline{{\rm MS}}}^{(6)}(\mu) \, .
\label{alphasDR}
\end{equation}
Actually, this relation summarizes three steps: first, the transition
from $g_{s,Y}$ in the $\overline{\rm MS}$~scheme to $g_s$ of the full MSSM in the $\overline{\rm DR}$scheme;
second, the decoupling of the SUSY particles, leading to $g_s$ running
with six (quark) flavor in the $\overline{\rm DR}$~scheme; third. the
transition to $g_{s,\overline{{\rm MS}}}^{(6)}$.
\eq{alphasDR} leads to the additional piece $C^{(2),\rm{shift \alpha_s }}_{fi}$ 
in \eq{Cdecomp}. 
\medskip

In principle we should have performed our calculation (of the full theory side) using dimensional reduction, 
which preserves supersymmetry, followed by modified minimal subtraction. 
The corresponding result for $C^{d\,LR\,(2)}_{fi}$ can be reconstructed by also
shifting the parameter $m_{\tilde{q}_s}$ and $m_{\tilde{g}}$ from the
$\overline{\rm MS}$~scheme to the $\overline{\rm DR}-$scheme in the
expression for $C^{d\,LR\,(1)}_{fi}$. As only $m_{\tilde{g}}$ gets such a
shift at the relevant order in $\alpha_s$, we denote this 
contribution in \eq{Cdecomp} as 
$C^{(2),m_{\tilde g_{ \overline{\rm MS}\,}}\to m_{\tilde g_{{\overline{\rm DR}}}}}_{fi}$. 
\medskip

This completes the derivation of the matching condition for
$C^{d\,LR\,(2)}_{fi}$ when ignoring the tadpole contribution (i.e. diagrams
1 and 2). Note that we performed our calculation using the
expression for the gluon propagator in an arbitrary $R_\xi$~gauge and
found a gauge-invariant result for $C^{d\,LR\,(2)}_{fi}$. 
\medskip

\begin{figure}
\centering
\includegraphics[width=0.8\textwidth]{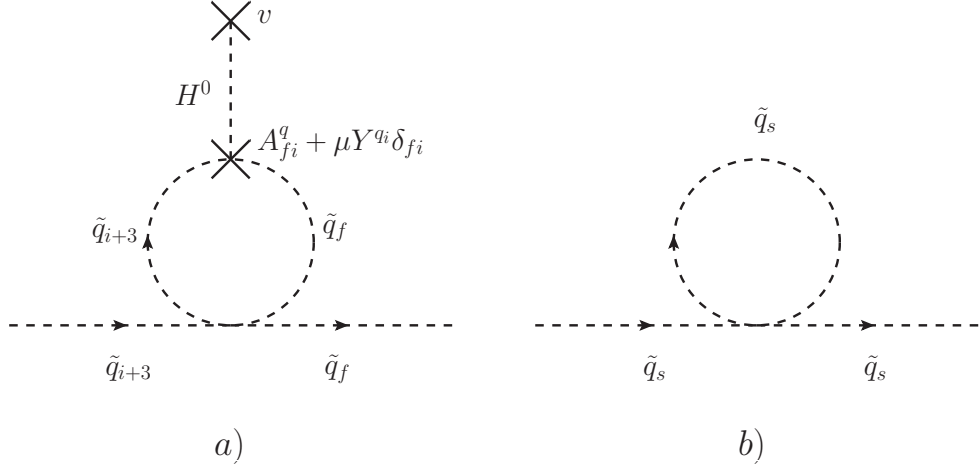}
\caption{Decomposition of the squark tadpole that is contained in
  diagram~2 of Fig.~\ref{2-loop-diagrams} as a subdiagram: In the
  decoupling limit the squark tadpole is either proportional to one
  element $\Delta^{q\,LR}_{ij}$ (a) or independent of
  $\Delta^{q\,LR}_{ij}$ (b). In the first case, it connects
  left-handed with right-handed squarks, while in the second case it
  is flavor and chirality conserving (proportional to
  $\delta_{st}$). The divergence of the piece proportional to
  $\Delta^{q\,LR}_{ij}$ is absorbed by the counterterms to $Y^q$ and
  $A^q$ while the divergence of the piece stemming from diagram~b) is
  canceled by a squark mass counterterm.
\label{squark-tadpole}}\hrule
\end{figure}

\subsection{The squark tadpole}
\label{sec:tadpole}

The diagrams containing a squark-tadpole self-energy as a subdiagram
require close examination. Diagram~1 vanishes but the squark-tadpole
contained in diagram~2 contains a divergence that enforces a
renormalization of both the physical squark masses and the trilinear
couplings of squarks to the Higgs field (the Yukawa couplings and the
$A$~terms). Thus it has to be decomposed into the corresponding two
parts. 
\medskip

Let us first consider the decoupling limit in which the expressions
are simpler but the structure of the divergences is the same as in the
full theory because higher powers (two or more) of
$\Delta^{q\,LR}_{ij}$ generate finite contributions only. In the
decoupling limit \eq{squark_squark_SE} simplifies to
\begin{equation}
-\frac{{\alpha_s}}{{4\pi }} \, {C_F} \left( {{\delta _{st}}m_{{{\tilde
        q}_s}}^2 - 2\sum\limits_{i,j = 1}^3 {\left(
    {\delta_{i^\prime+3,s}\Gamma^{i i^\prime \star}_{QR}\Delta
      _{ij}^{q\,RL}\Gamma^{jj^\prime}_{QL}\delta_{j^\prime t} +
      \delta_{i^\prime s}\Gamma^{ii^\prime \star}_{QL}\Delta
      _{ij}^{q\,LR}\Gamma^{j j^\prime}_{QR}}\delta_{j^\prime+3, t}
    \right)} } \right)\frac{1}{\varepsilon } + {\rm finite}\,.
\label{squark-tadpole-decoupling}
\end{equation}
Here we clearly see that to render the first term in
\eq{squark-tadpole-decoupling} finite, which is flavor diagonal
(corresponding to Fig.~\ref{squark-tadpole} (b)),
a renormalization of the squark masses is necessary. On the other hand, for canceling  the divergence of the second term in
\eq{squark-tadpole-decoupling} (corresponding to diagram a) in
Fig.~\ref{squark-tadpole}), which is proportional to $\Delta^{q\,LR}_{ij}$,
 a counterterm to the Yukawa
coupling and the $A$~term contained in $\Delta^{q\,LR}_{ij}$ is necessary. 
The latter point can be seen as follows:
In the decoupling limit the amputated chirality-changing squark
two-point function for $\tilde{q}_{j'}^L \to \tilde{q}_{i'}^R$ is
given, at lowest order in $\alpha_s$, by
\begin{equation}
\Gamma^{i i^\prime
  \star}_{QR}\Delta_{ij}^{q\,RL}\Gamma^{jj^\prime}_{QL} \, .
\end{equation}
From this we can read off the common renormalization renormalization
constant $Z_Y$ of the Yukawa couplings $Y^{\tilde{q}_i}$ and the
$A^q_{ij}$ and the $A^{'q}_{ij}$ terms, obtaining in the minimal
subtraction scheme ($\overline{\rm DR}$ or $\overline{\rm MS}$)
\begin{equation}
Z_{Y} = 1- \dfrac{\alpha_s}{4\pi} \, 2 \, C_F \, \dfrac{1}{\varepsilon} \, .
\end{equation}
In fact, it turns out that this renormalization of the Yukawa
couplings is necessary for maintaining supersymmetry
with respect to the Yukawa
coupling involved quark-quark-Higgs coupling and the one of the 
squark-squark-Higgs coupling.
\medskip

\subsection{Result for $C^{d\;LR}_{fi}$ retaining all powers of  $v/M_{\rm SUSY}$}
\label{sec:Exact_results}
For the Wilson coefficient $C_{fi}^{d\;LR}$ of the two-quark operator $\overline{q}_f P_R q_i$ we write the general decomposition 
\begin{equation}
C_{fi}^{d\,LR}=C_{fi}^{d\,LR(1)}+C_{fi}^{d\,LR(2)}\equiv  \frac{\alpha_s}{4\pi} C^{(1)}_{fi} + \left( \frac{\alpha_s}{4\pi}
\right)^2 C^{(2)}_{fi}\,.
\end{equation}
From \eq{SigmaLR1} we directly obtain
\begin{eqnarray}
C^{(1)}_{fi} = \sum\limits_{t=1}^6 \left( 4 \, m_{\tilde{g}} \, 
 C_F \, W_{ft}^{\tilde d} W_{i + 3,t}^{\tilde d\star}  \dfrac{{x_t^2 \ln \left( {x_t }
    \right)}}{{1 - x_t^2 }} \right) \,.
\label{Cleading}
\end{eqnarray}
Here we introduced the abbreviations 
\begin{equation}
x_t=m_{\tilde{d}_t}/m_{\tilde{g}}\,,
\end{equation}
and for later convenience we also define
\begin{equation}
y_t=m_{\tilde{u}_t}/m_{\tilde{g}},, \qquad \,x_\mu={\mu}/m_{\tilde{g}}\,, 
\end{equation}
where $\mu$ is the renormalization scale.
\medskip

According to the detailed description in the previous subsections, we
decompose the Wilson coefficient 
$C^{(2)}_{fi}$ into various pieces:
\begin{equation}
C^{(2)}_{fi}=C^{(2),1}_{fi} +
        C^{(2),2}_{fi} +
          C^{(2),3}_{fi} +
          C^{(2),4}_{fi} +
          C^{(2),\mu}_{fi}+
          C^{(2),\rm{shift \alpha_s }}_{fi} +  
          C^{(2),m_{\tilde g_{ \overline{\rm MS}\,}}\to m_{\tilde g_{{\overline{\rm DR}}}}}_{fi} +
          C^{(2),TP}_{fi} \, .
\label{Cdecomp}
\end{equation}

%
%
\noindent We freely made use of the unitarity of the mixing matrices $W^{\tilde{q}}$ and obtain
\begin{equation}
\renewcommand{\arraystretch}{1.8}
\begin{array}{l}
 C_{fi}^{(2),1}  = \sum\limits_{j=1}^{3} \, \sum\limits_{s,t=1}^{6} \left\{
 2 \, W_{ft}^{\tilde d} \, 
 W_{j+3,t}^{\tilde d \star} \,
 W_{i+3,s}^{\tilde d \star} \,
 W_{j+3,s}^{\tilde d} \,
 m_{\tilde{g}} \,C_F \, \left(2\,C_F  - C_A \right) 
\, \dfrac{1}{(1-x_s^2) \, (1-x_t^2)}  \right. \\ 
\qquad  \times \left[ \left(1 - x_s^2\right)^2 \, {\rm Li}_2 \left(1 -x_s^2\right) 
- \left(1 - x_t^2\right)^2 \, {\rm Li}_2 \left(1 - x_t^2\right) + 
\left(x_s^2  - x_t^2\right)^2 \, {\rm Li}_2 \left(1 - x_t^2 /x_s^2\right)
 \right. \\ 
\qquad\;\;\;\;\;-4 \, x_t^2 \left( x_t^2-x_s^2 \right)\ln(x_s) \ln(x_t)
+ 6 \, x_s^2 \left( x_t^2-1 \right) \ln(x_s)
- 6 \, x_t^2 \left( x_s^2-1 \right) \ln(x_t) \\
\qquad\;\;\;\;\; \left. \left. + 2 \, x_t^2 \left( x_s^2-1 \right) \ln^2(x_t)
+ 2 \, \left( x_s^4 + x_t^4 - 3 \,x_t^2 \, x_s^2 + x_s^2 \right) \, \ln^2(x_s)
\right] \right\} \\
\qquad\;\;\;\;\; + \sum\limits_{t=1}^{6} \left\{
 4 \, W_{ft}^{\tilde d} \, 
 W_{i+3,t}^{\tilde d \star} \,
 m_{\tilde{g}} \,C_F \, \left(2\,C_F  - C_A \right) \, \dfrac{x_t^2}{\left( 1-x_t^2 \right)^2}
 \right. \\
\qquad \;\;\;\;\;\left. \times \left[
\left(1-2 \, x_t^2 \right) \, \ln^2(x_t)-2  \, 
\left(1-x_t^2\right) \, \ln(x_t) \right] \right\}\,,
\end{array} 
 \label{part1}
\end{equation}
%

\begin{equation}
\renewcommand{\arraystretch}{1.8}
\begin{array}{l}
 C_{fi}^{(2),2}  = \sum\limits_{s,t=1}^6 \left\{ \dfrac{{W_{ft}^{\tilde d} \, W_{i+3,t}^{\tilde d \star} 
\, tr \, m_{\tilde{g}} \,C_F }}{{\left( {1 - x_t^2 } \right)^2  }} \right. \\ 
  \times \left\{ {4 {\left( {1 - x_s^2 } \right)\left( { - x_s ^2  + 2\,x_t ^2  - 1} \right){\rm{Li}}_2 (1 - x_s^2 )}} \right. \\ 
 \;\;\;\; - 4 {\left( {x_s  + x_t } \right)^2 \left( {x_s  - x_t } \right)^2 {\rm{Li}}_2 (1 - x_s^2 /x_t^2 )}  \\ 
 
 \;\;\;\; - 4 x_t^2 x_s^2 \left(1+\left(4-2x_s^2\right)\ln\left(x_s\right)\right)\left(1-x_t^2+\left(1+x_t^2\right)\ln\left(x_t\right)\right)\\
    {\;\;\;\; - \dfrac{1}{3}\left[ {48\left( {\ln\! \left( {x_t } \right)\ln\! \left( {x_s } \right)x_s^2 \left( {x_t^2  - x_s^2  - x_t^4 } \right) + \ln\!^2\! \left( {x_t } \right)x_t^2 \left( {x_t^2  - x_s^2 } \right) - \ln\! \left( {x_t } \right)x_t^4  - x_t^2 } \right)} \right.} \hfill  \\
   {\;\;\;\;\;\;\;\;\;\;\; + 24\left( {\ln\! \left( {x_t } \right)\ln\! \left( {x_s } \right)x_s^4 x_t^2 \left( {1 + x_t^2 } \right) + \ln\! \left( {x_s } \right)x_s^2 x_t^2 \left( {x_t^2  - 1} \right) + \ln\! \left( {x_t } \right)x_t^2  + \ln\!^2\! \left( {x_t } \right)x_s^4 } \right)} \hfill  \\
   {\;\;\;\;\;\;\;\;\;\;\; + 12\left( {\ln\! \left( {x_t } \right)x_s^2 x_t^2 \left( {1 - x_t^2 } \right) + \ln\! \left( {x_s } \right)x_s^4 \left( {1 - x_t^4 } \right) - x_s^2 x_t^2 } \right)} \hfill  \\
   \left. {\left. {\left. {\;\;\;\;\;\;\;\;\;\;\; + 6x_s^2  + 6x_s^2 x_t^4  +
         30x_t^4 + 18}
       \right]} \right\}} \right\} \,,\hfill  \\
 \end{array}
  \label{part2}
\end{equation}

\begin{eqnarray}
&&C^{(2),3}_{fi} = C^{(2),2}_{fi}(x_s \to y_s)\,,
 \label{part3}
\end{eqnarray}

\begin{equation}
\renewcommand{\arraystretch}{1.8}
\begin{array}{l}
 C_{fi}^{(2),4}  = \sum\limits_{t=1}^{6} \left\{ 2\, W_{f,t}^{\tilde d} \, W_{i+3,t}^{\tilde d \star}
 \, m_{\tilde{g}} \,C_F  \right. \\ 
  \times \left\{ {\left( { - 3\,C_A  + 2\,C_F } \right){\rm{Li}}_2 (1 - x_t^2 )}  
  \right.  \\ 
 \;\;\;\;\; + \dfrac{1}{{3\left( {1 - x_t } \right)^2 \left( {1 + x_t } \right)^2 }} \\ 
 \;\;\;\;\; \times \left[ {\rm{tr}} \,n_f \left( (24 \ln^2(x_t)-12) \,
   x_t^4 +(24 \ln(x_t) +12) \, x_t^2 \right)
 \right.\\ 
 \;\;\;\;\;\;\;\;\; + 6 \, C_A \left(
   (3 \ln^2(x_t)-11 \ln(x_t)+9) \, x_t^4 +(3 \ln(x_t) - 14) \, x_t^2 +5 \right) \\ 
 \left. \left. \left. \;\;\;\;\;\;\;\;\; + 3 \, C_F \left( - (2
     \ln(x_t)+1) \, x_t^4 -(12 \ln^2(x_t) -12 \ln(x_t) +7) \, x_t^2 +
     8 \ln(x_t) +8 \right)
   \right] \right\} \right\} \,,\\ 
 \label{part4}
 \end{array}
\end{equation}

\begin{equation}
\renewcommand{\arraystretch}{1.8}
\begin{array}{l}
 C_{fi}^{(2),\mu}  = \sum\limits_{t=1}^{6} \left\{ -4 \, W_{f,t}^{\tilde d} \, W_{i+3,t}^{\tilde d \star}
 \, m_{\tilde{g}} \,C_F  \, \dfrac{\ln(x^2_{\mu})}{(1-x_t^2)^2} \right. \\ 
 \;\;\;\;\; \times \left[ {\rm{tr}} \,n_f \left( 
-2 x_t^2 \, ((2 \ln(x_t)-1) \, x_t^2 + 1) \right)
 \right.\\ 
 \;\;\;\;\;\;\;\;\; + C_A \left(
3 x_t^2 \,    ((2 \ln(x_t)-1) \, x_t^2 + 1) \right) \\ 
 \left. \left. \;\;\;\;\;\;\;\;\; + \dfrac{C_F}{2} \left(
-(4\ln(x_t)-1) x_t^4 + (2 \ln(x_t)+3) x_t^2 - 8 \ln(x_t) -4
\right)
   \right] \right\} \, , \\ 
 \label{partmu}
 \end{array}
\end{equation}
%

\begin{equation}
\renewcommand{\arraystretch}{2.4}
\begin{array}{l}
C^{(2),\rm{shift \alpha_s }}_{fi} = 
\sum\limits_{t=1}^6\left\{-\dfrac{4 \, W_{f,t}^{\tilde d} \, W_{i+3,t}^{\tilde d \star} \, m_{\tilde{g}} \, C_F\, x_t^2 \, \ln(x_t)}{3 \,
  (1-x_t^2) } \, \left[\; \sum\limits_{s=1}^{6}( \ln(x_s) + \ln(y_s)) -4 C_A +
  3 C_F \right] \right. \\ \qquad\qquad\!\!\qquad\qquad\left.
+\dfrac{4 \left(n_f+6 \right) \, W_{f,t}^{\tilde d} \, W_{i+3,t}^{\tilde d \star} \,  m_{\tilde{g}} \, C_F \, x_t^2 \, \ln(x_t)}{
 3\, (1-x_t^2)} \, \ln(x_\mu^2)\right\}\,,
  \end{array}
  \label{partshift}
\end{equation}

\begin{equation}
C^{(2),m_{\tilde g_{ \overline{\rm MS}\,}}\to m_{\tilde g_{{ \overline{\rm DR}}}}}_{fi}=- 2 \, \sum\limits_{t=1}^6 \left\{ W_{ft}^{\tilde d} \,
W_{i+3,t}^{\tilde d \star} \, m_{\tilde{g}} \, C_F \, 
  \, C_A \, \dfrac{\left( 1+x_t^2 \right) \, \left( 1-x_t^2+ 2 x_t^2
    \, \ln(x_t)\right)}{(1-x_t^2)^2} \right\} \, ,
    \label{shiftDRgluino}
\end{equation}

\begin{eqnarray}
\renewcommand{\arraystretch}{2.4}
\begin{array}{c}
  C^{(2),TP}_{fi} = -2 \, m_{\tilde g} C_F^2 \sum\limits_{t = 1}^6
  \left\{ W_{ft}^{\tilde d} \, W_{i + 3,t}^{\tilde d\star}
\dfrac{x_t^2}{(1-x_t^2)^2} \, \left( 1-x_t^2+2\ln(x_t) \right) \,
\left( 1-2\ln(x_t)+ \ln(x^2_{\mu})\right) \right\}  \\
-8 \, m_{\tilde g} C_F^2 \sum\limits_{j,j' = 1}^3
  {\sum\limits_{s,t,s' = 1}^6 {\left[ {W_{fs}^{\tilde d} \left( {W_{j' + 3,s}^{\tilde d\star} W_{j' + 3,t}^{\tilde d} W_{jt}^{\tilde d\star} W_{js'}^{\tilde d}  + W_{j's}^{\tilde d\star} W_{j't}^{\tilde d} W_{j + 3,t}^{\tilde d\star} W_{j + 3,s'}^{\tilde d} } \right)W_{i + 3,s'}^{\tilde d\star} } \right.} }  \\ 
 \left. { \times \dfrac{{x_{t}^2 \left( {2 \, \ln \left( {x_{t} }
         \right) - \ln \!\left( x_\mu^2 \right) - 1} \right)\left(
       {x_s^2 x_{s'}^2 \ln \!\left( {\dfrac{{x_{s'} }}{{x_s }}}
         \right) + x_{s}^2 \ln \left( {x_s } \right) - x_{s'}^2 \ln
         \left( {x_{s'} } \right)} \right)}}{{\left( {x_s^2  -
         x_{s'}^2 } \right)\left( {x_s^2  - 1} \right)\left( {x_{s'}^2
         - 1} \right)}}} \right] \,.\\ 

 \end{array}
   \label{partTP}
\end{eqnarray}
\noindent
In the MSSM we have
\begin{equation}
C_A=3\,,\qquad C_F=4/3\,,\qquad tr=1/2 \qquad {\rm and}\qquad n_f=6 \,.
\end{equation}
To summarize, Eqs.~(\ref{Cleading}) and (\ref{Cdecomp}) contain the
full result for the Wilson coefficient $C^{d\;LR}_{fi}$ where
the $A$~terms, the Yukawa coupling, the squark and the gluino masses
of the MSSM 
are renormalized in the $\overline{\rm{DR}}$ scheme, while $g_s$ stands for the strong coupling constant of the SM in the 
$\overline{{\rm MS}}$ scheme, running with six flavors. The effective
operators, or equivalently the Wilson coefficients, are understood to be renormalized according to the $\overline{{\rm MS}}$ scheme.
\medskip

So far, we discussed the derivations of $C^{d\;LR}_{fi}$. 
The corresponding result $C^{d\;LR}_{fi}$ for up quarks can be
obtained by replacing $W_{}^{\tilde d}$ with $W_{}^{\tilde u}$ and exchanging $x$ and $y$.

\subsection{Reduction of the matching scale dependence at NLO}
\label{MatchingReduction}

The purpose of our NLO calculation is also the reduction of the
matching scale dependence of the effective Higgs couplings that can serve as an estimate of the theory uncertainty. This
reduction not only is an improvement achieved by our NLO
calculation but also serves as an additional check of its correctness. 
\medskip

As we will see in the next section, the quantity directly related to the
Higgs couplings is $\hat C_{fi}^{q\,LR}$ defined as
\begin{equation}
\renewcommand{\arraystretch}{2.0}
\begin{array}{l}
 \hat C_{fi}^{q\,LR} = C_{fi}^{q\,LR\,(1)}+C_{fi}^{q\,LR\,(2)}  + \dfrac{1}{2}\left( {C_{ff}^{q\,LL} C_{fi}^{q\,LR\,(1)}  + C_{fi}^{q\,LR\,(1)} C_{ii}^{q\,LL} } \right) +O\left(\alpha_s^2,\,\alpha_s^3\tan\beta\right) \,.
 \end{array}
 \label{SigmaHut}
\end{equation}
We use in the following the decomposition $\hat C_{fi}^{q\,LR}=\hat
C_{fi}^{q\,LR,(1)}+\hat C_{fi}^{q\,LR,(2)}$. At LO in our counting of $\alpha_s$ and $\tan\beta$ we have $\hat C_{fi}^{q\,LR,(1)}=C_{fi}^{q\,LR,(1)}$.
\medskip

$\hat C _{fi}^{q\,LR}$ (and thus also $\hat C_{fi}^{q\,LR\,(1)}$) at a fixed low
scale $\mu _{\rm{low}}$ is obtained from $\hat C _{fi}^{q\,LR}$ at the matching scale $\mu_{0}$ via
\begin{equation}
\hat C^{q\,LR}_{ij}  \left( {\mu _{\rm{low}} } \right) = U\left( {\mu _{\rm{low}} ,\mu _{0} } \right)\hat C^{q\,LR}_{ij}  \left( {\mu _{0} } \right) \,.
\label{Sigma_low_high}
\end{equation}
This evolution is the same as for the quark masses in the SM. The explicit NLL
expression can be taken, e.g., from Eq.~(4.81) in Ref.~\cite{Buras:1998raa}. 
It is this expression that we use for the numerical study in
Sec.~\ref{numerics} when doing the evolution to the low scale $\mu_{\rm low}$.
\medskip

However, for showing analytically the reduced matching scale
dependence, it is sufficient to assume that the scale $\mu_{\rm low}$ is
close to the matching scale $\mu_0$ so that it is not necessary to
resum large logarithms. In this case the evolution matrix 
$U\left( {\mu _{\rm{low}} ,\mu _{0} }\right)$ can be expanded as
\begin{equation}
U\left( {\mu _{low} ,\mu _{0} } \right) \approx 1 + \alpha _s \left(
{\mu _{0} } \right)\frac{{\gamma _m^{\left( 0 \right)} }}{{8\pi }}\ln
\left( {\frac{{\mu _{0}^2 }}{{\mu _{\rm low}^2 }}} \right),\;\;\;\;\gamma
_m^{\left( 0 \right)}  = 6 \, C_F \,.
\label{U_low_high}
\end{equation}
At LO $\hat{C}^{q\,LR}_{ij}$ depends only implicitly on the
renormalization scale via the scale dependence of various parameters.
For small changes of the original matching scale $\mu_{0}$ to a new matching scale $\mu$, we get
\begin{equation}
\frac{{\hat C^{q\,LR(1)}_{ij} \left( \mu  \right)}}{{\hat C^{q\,LR(1)}_{ij}
    \left( {\mu _{0} } \right)}} 
\approx 1 + \dfrac{\alpha _s \left(\mu _{0} \right)}{4\pi} \left(\beta_0 + S \right) \ln \left( {\frac{{\mu _{0}^2 }}{{\mu ^2 }}} \right) \,.
\end{equation}
The contribution involving $\beta_0$  comes from
expressing $\alpha_s(\mu)$ in terms of $\alpha_s(\mu_0)$, while the
one involving $S$ is attributable to the corresponding manipulation of the
squark and gluino masses, the Yukawa couplings and the $A$ (and $A'$) terms.   
Together with \eq{Sigma_low_high} and \eq{U_low_high} the variation of the matching scale leads to the following ratio
\begin{equation}
\frac{{U\left( {\mu _{low} ,\mu } \right) \hat C^{q\,LR(1)}_{ij} \left(
    \mu  \right)}}{{U\left( {\mu _{low} ,\mu_{0} }
    \right) \hat C^{q\,LR(1)}_{ij} \left( {\mu _{0} } \right)}} \approx 
1 + \dfrac{\alpha _s \left( {\mu _{0} } \right)}{4\pi}
\left( \beta_0  + S - \frac{\gamma_m^{\left( 0 \right)} }{2} \right)\ln \left( {\frac{{\mu _{0}^2 }}{{\mu ^2 }}} \right)\,.
\end{equation}
The explicit $\mu$ dependence proportional to $\alpha_s$ in this ratio 
has to be compensated when going to NLO.

\medskip

The piece of $\hat C^{q\,LR(2)}_{ij} $ with explicit scale dependence 
(with contributions from Eqs.~(\ref{partmu}), (\ref{partshift}), (\ref{partTP}), and from
\eq{SigmaHut} through $C^{q\;LL}_{ff}$ and  $C^{q\;LL}_{ii}$), 
can be compactly written as 
\begin{equation}
\hat C^{q\,LR\,(2),\rm{\mu}}_{ij}(\mu) =
\dfrac{\alpha_s(\mu)}{4\pi} \, \left[  
S - 2 \, {\rm tr} \, n_f  +
3  \, C_A  -3  \, C_F  +  \dfrac{n_f}{3} + 2  \right] \,
\hat C^{q\,LR\,(1)}_{ij}(\mu) \, \ln(x^2_{\mu}) \, .
\end{equation}
Using this information, we finally get at NLO
\begin{equation}
\renewcommand{\arraystretch}{2.0}
\begin{array}{l}
\dfrac{{U\left( {\mu _{low} ,\mu } \right)\hat{C}^{q\,LR}_{ij} \!\left(
    \mu  \right)}}{{U\left( {\mu _{low} ,\mu_{0} }
    \right)\hat{C}^{q\,LR}_{ij} \!\left( {\mu _{0} } \right)}} \\ \approx
1+ \dfrac{{\alpha _s \left( {\mu _{0} } \right)}}{{4\pi }}\left(
\beta_0 + S - \dfrac{\gamma _m^{\left( 0 \right)}}{2}  - S + 
2 \, {\rm tr} \, n_f  -
3 \, C_A  +3  \, C_F -  \dfrac{n_f}{3} - 2 \! \right) \!
\ln \left( {\dfrac{{\mu _{0}^2 }}{{\mu ^2 }}} \right) = 1  \,,
\end{array}
\label{mu_cancelation}
\end{equation}
as expected. 

\subsection{Numerics}
\label{numerics}

\begin{figure}
\centering
\includegraphics[width=0.7\textwidth]{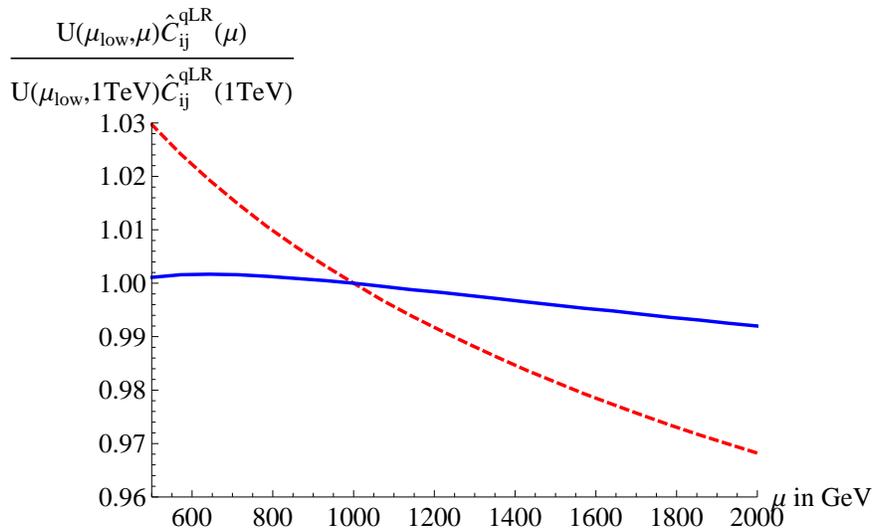}
\caption{Dependence on the matching scale $\mu$ of the one-loop and
  two-loop results for $\hat C _{fi}^{q\,LR}(\mu_{\rm low})$, using $M_{\rm SUSY}=1$~TeV and $\mu_{\rm
    low}=m_W$. Red (dashed): matching done at LO;  blue (darkest):
  matching done at NLO matching.  As expected, the
  matching scale dependence is significantly reduced. For the one-loop result, $\hat
  C _{fi}^{q\,LR}$ is understood to be $C_{fi}^{q\,LR\,(1)}$ (see text).
  \label{mu-abhaengigkeit}}\hrule
\end{figure}

\begin{figure}
\centering
\includegraphics[width=0.7\textwidth]{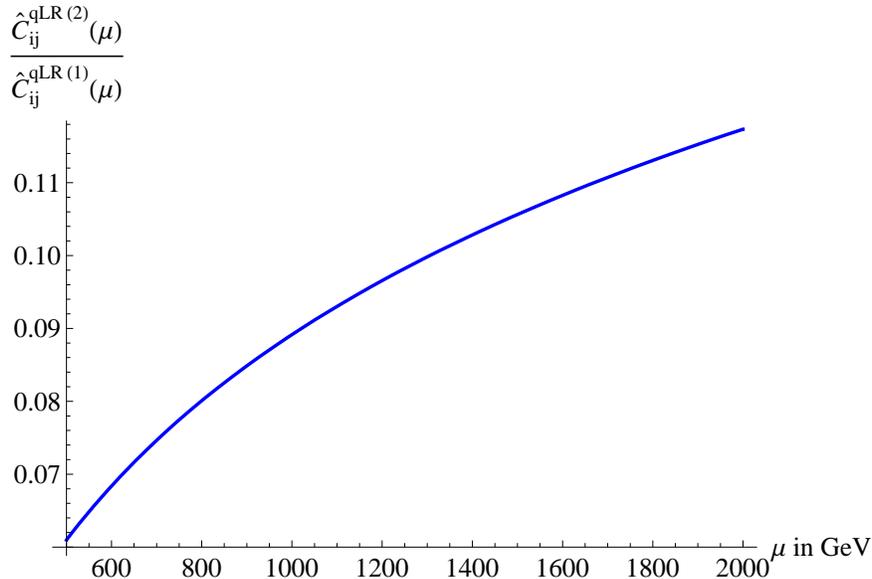}
\caption{Relative importance of the two-loop corrections as a function of the matching scale $\mu$. 
We see that the two-loop contribution is approximately +9\% of the
one-loop contribution for $\mu=M_{\rm SUSY}=1\, {\rm TeV}$. $\hat C _{fi}^{q\,LR}$
is defined in \eq{SigmaHut}.
  \label{2loop-1loop}}\hrule
\end{figure}

In this section we study the numerical importance of our two-loop corrections and the reduced matching scale dependence compared to the one-loop result.
\medskip

The matching scale dependence, as shown in Fig.~\ref{mu-abhaengigkeit}
for SUSY masses of 1 TeV, is significantly reduced as expected from the
previous subsection. Note that the
relative importance of the NLO result is to a very good approximation
independent of the size of $\Delta^{q\,LR}_{ij}$.
\medskip

The relative importance of the two-loop contribution to $\hat{C}^{q\;LR}_{ij}(\mu)$ is shown in
Fig.~\ref{2loop-1loop} as a function of the matching scale $\mu$. 
For SUSY masses of 1 TeV the $\alpha_s^2$
corrections lead to a constructive contribution of approximately
$9 \%$ compared to the one-loop result 
that is in agreement with Ref.~\cite{Noth:2010jy}. Again, the
relative importance of the NLO result is to a very good approximation
independent of the size of $\Delta^{q\,LR}_{ij}$.

\subsection{Transition to the decoupling limit}
\label{decouplinglimit}

While the two-loop contributions calculated in this section are
obtained in the approximation $p=m_q=0$, the results given in
Sec.~\ref{sec:Exact_results}
still contain all powers $v/M_{\rm{SUSY}}$ implicitly via the squark
mixing matrices $W^{\tilde{q}}$ and the physical squark masses $m_{\tilde{q}_s}$ involved. The
transition to the decoupling limit, in which all chirally enhanced
corrections can be resummed analytically, can be done by the following
prescription. 

\medskip

In all parts of the genuine two-loop contributions listed above
(\eq{part1}--\eq{shiftDRgluino}) only two mixing
matrices occur, except in \eq{part1} and \eq{partTP}. \eq{part1} contains the
following combinations of mixing matrices and a loop-function $f$
which depends on squarks masses $m_{\tilde q_s}$ and $m_{\tilde q_t}$ 
\begin{equation}
\sum\limits_{s,t = 1}^6\sum\limits_{j = 1}^3 W_{ft}^{\tilde d} \, 
W_{j+3,t}^{\tilde d \star} \,
W_{i+3,s}^{\tilde d \star} \,
W_{j+3,s}^{\tilde d}\, f(x_s^2,\,x_t^2)\,.
\label{WWWW}
\end{equation}
Note that in the decoupling limit, the squark with index $s$ in \eq{WWWW} must be a linear combination of right-handed squark only, since otherwise at least two chirality changes (two insertions of $\Delta^{d\,LR}_{ij}$) would be necessary. Thus we can replace 
\begin{equation}
 W_{i+3,s}^{\tilde d \star} \,W_{j+3,s}^{\tilde
   d}\,\to\;\Gamma^{ik\star}_{DR}\Gamma^{jk}_{DR}=\Lambda^{d\,RR}_{k\,ji}\qquad{\rm
   and}\qquad x_s^2\;\to\;x_{Rk}^2 \, ,
 \label{WtoGamma}
\end{equation}
where $k$ only runs from 1 to 3 and we defined 
\begin{equation}
x_{L(R)k}^2=\dfrac{m^2_{\tilde q_{k}^{L(R)}}}{m_{\tilde g}^2}	\,.
\end{equation}
The resulting expression
\begin{equation}
\sum\limits_{t = 1}^6 \sum\limits_{k,j = 1}^3 W_{ft}^{\tilde d} \, 
W_{j+3,t}^{\tilde d \star} \,
\Lambda^{d\,RR}_{k\,ji}\, f(x_{Rk}^2,\,x_t^2)
\label{WWLambda}
\end{equation}
can now be expanded in powers of $v/M_{\rm SUSY}$ which amounts at leading order to the replacement
\begin{equation}
\sum\limits_{t = 1}^6 W_{f t}^{\tilde d} W_{j+ 3,t}^{\tilde d \star}
f(\ldots \, ,x_t^2)\;\to\;\sum\limits_{m,n,j^\prime, j^{\prime\prime} =
  1}^3\Lambda_{m\;f j^{\prime\prime}}^{d\,LL} \Delta
_{j^{\prime\prime}j^\prime}^{d\,LR} \Lambda_{n\;j^\prime
  j}^{d\,RR}\dfrac{f(... \, , x_{Lm}^2)-f(... \, , x_{Rn}^2)}{m^2_{\tilde
    q_{m}^L}-m^2_{\tilde q_{n}^R}}\,,
\label{ExpansionDeltaLR}
\end{equation}
where the dots represent possible additional dependences on squark masses. Now we apply \eq{ExpansionDeltaLR} to \eq{WWLambda} and use 
\begin{equation}
\sum\limits_{j=1}^3	\Lambda _{m\;fj}^{d\,(LL)RR}\Lambda
_{n\;ji}^{d\,(LL)RR} =\Lambda _{m\;fi}^{d\,(LL)RR} \delta_{mn} \, .
	\label{LambdaLambda}
\end{equation}
The final result for \eq{WWWW} in the decoupling limit is then 
\begin{equation}
\sum\limits_{m,n,j^\prime,j^{\prime\prime}=1}^3	\Lambda _{m\;f
  j^{\prime\prime}}^{d\,LL} \Delta _{j^{\prime\prime}j^\prime}^{d\,LR}
\Lambda _{n\;j^\prime i}^{d\,RR} \,
\dfrac{f(x_{Rn}^2,x_{Lm}^2)-f(x_{Rn}^2,x_{Rn}^2)}{m^2_{\tilde q_{m}^L}-m^2_{\tilde q_{n}^R}}\,.
\label{DecouplingLimit}
\end{equation}

For \eq{partTP} a similar procedure works. It contains the following
combination of mixing matrices with a loop function depending on three
different squark masses with the indices $s$, $t$, and $s^\prime$ 
\begin{equation}
\sum\limits_{s,t,s^\prime=1}^6\sum\limits_{j,j^\prime=1}^3
           {W_{fs}^{\tilde d} \left( {W_{j' + 3,s}^{\tilde d\star}
               W_{j' + 3,t}^{\tilde d} W_{jt}^{\tilde d\star}
               W_{js'}^{\tilde d}  + W_{j's}^{\tilde d\star}
               W_{j't}^{\tilde d} W_{j + 3,t}^{\tilde d\star} W_{j +
                 3,s'}^{\tilde d} } \right)W_{i + 3,s'}^{\tilde
               d\star} }\, f(x_s^2,x_t^2,x_{s^\prime}^2)\,.
	\label{WWWWWW}
\end{equation}
Note that the first term in \eq{WWWWWW} vanishes in the decoupling
limit since it necessarily involves multiple chirality flips. For the
second term two replacements analogous to \eq{WtoGamma} have to be
performed, and after using two times the relation in \eq{LambdaLambda}
the decoupling limit of \eq{WWWWWW} reads
\begin{equation}
\sum\limits_{m,n,j^\prime,j^{\prime\prime}=1}^3 \!\!\!	\Lambda _{m\;f
  j^{\prime\prime}}^{d\,LL} \Delta _{j^{\prime\prime}j^\prime}^{d\,LR}
\Lambda _{n\;j^\prime i}^{d\,RR}\,
\dfrac{f(x_{Lm}^2,x_{Lm}^2,x_{Rn}^2)-f(x_{Lm}^2,x_{Rn}^2,x_{Rn}^2)}{m^2_{\tilde
    q_{m}^L}-m^2_{\tilde q_{n}^R}}\,.
\end{equation}
This result involves the same combination of mixing
matrices as the one in \eq{DecouplingLimit}. 
\medskip

To all other parts of $C_{fi}$ the rule in \eq{ExpansionDeltaLR} can be applied directly to obtain the corresponding expression in the decoupling limit.
\medskip

\section{Relations between quark masses and the MSSM Yukawa couplings at NLO}
\label{RenReg}

Beyond one-loop \eq{mb_Yb_gen} and \eq{md-Yd} for the determination of $Y^d$ can easily be generalized to higher loop orders because the chirality changing self-energy (and also the resulting Wilson coefficient) is still proportional to one element $\Delta^{d\;LR}_{ij}$ in the decoupling limit, as shown in Sec.~\ref{decouplinglimit}. However, since we are dealing with order one corrections, we must specify how we count contributions at higher loop orders in $\alpha_s$. $C_{fi}^{q\,LR\,\left( 1 \right)}$ is proportional $\alpha_s\tan\beta$ and $C_{fi}^{q\;LR\;\left( 2 \right)}$ is proportional to $\alpha_s^2\tan\beta$. Here, $\tan\beta$ stands schematically for a chiral enhancement factor, also including $A^q_{ij}/(Y^q_{ij}\Msusy)$. We will count $\alpha_s \tan\beta$ as order one and thus $\alpha_s^2 \tan\beta$ as order $\alpha_s$. Since $C_{fi}^{q\,LL,RR}$ is not chirally enhanced, the only relevant term in our approximation (of order $\alpha_s$) is the one-loop contribution. Thus, $C_{fi}^{q\,LL,RR}$ is always understood to be the one-loop contribution proportional to $\alpha_s$. 
\medskip

To derive the relation between the quark masses and the Yukawa couplings of the MSSM superpotential at NLO we also need to specify the renormalization scheme used for the matching procedure. Let us explicitly denote the renormalization scheme for the quantities in the matching condition \eq{mb_Yb_gen} (at the scale $m_{\rm SUSY}$) which is important at NLO:
\begin{equation}
v_q Y^{q_i\,\overline{\rm MS}}_{\rm{eff}}=m_{q_i}^{\overline{\rm MS}} = \left(v_q Y^{q_i\,\overline{\rm MS}}_{\rm tree}  + C_{ii\,\overline{\rm MS}}^{q\,LR\,(1)}+C_{ii}^{q\,LR\,(2)}\right)\times\left(1+ \dfrac{1}{2} \left( {C_{ii}^{q\,LL}  + C_{ii}^{q\,RR} } \right)\right)  \,. \label{mb_Yb_gen_NLO}
\end{equation}
Again, $Y^{q_i\,\overline{\rm MS}}_{\rm tree}$ is the Wilson coefficient induced via the Yukawa coupling of the MSSM. This means at the matching scale it is given by:
\begin{equation}
Y^{q_i\,\overline{\rm MS}}_{\rm tree}\left(\mu_{\rm SUSY}\right)=Y^{q_i}_{\overline{\rm MS}}\left(\mu_{\rm SUSY}\right)=\left(1+\frac{\alpha_s}{4\pi}C_F\right)Y^{q_i}_{\overline{{\rm DR}}}\left(\mu_{\rm SUSY}\right)\,.
\label{YMSDR}
\end{equation}
In  our counting in $\alpha_s$ and $\tan\beta$ the renormalization scheme for $C_{ii}^{q\,LL}$ and $C_{ii}^{q\,RR}$ is irrelevant. Note that the quark mass $m_{q_i}$ is understood to be evaluated at the matching scale. Further, one should recall from the last section that despite the fact that we renormalized $C^{q\,LR}_{ii}$ in the $\overline{\rm MS}$ scheme, it contains parameters given in the $\overline{\rm DR}$ scheme, e.g. $Y^{\tilde q_i}=Y^{q_i}_{\overline{\rm DR}}$. Since we are interested in $Y^{\tilde q_i}$, the Yukawa coupling of the MSSM superpotential, we must express $Y^{q_i\,\overline{\rm MS}}_{\rm tree}$ in \eq{mb_Yb_gen_NLO} in terms of $Y^{\tilde q_i}_{\overline{\rm DR}}$ via \eq{YMSDR} so that we can solve for $Y^{q_i}_{\overline{\rm DR}}$.
\medskip

In conclusion we arrive at the NLO generalization (order $\alpha_s^2\tan\beta$) of \eq{mb_Yb_gen}:
\begin{equation}
Y^{d_i}_{\overline{{\rm DR}}} =\dfrac{m_{d_i}^{\overline{{\rm MS}}} - \hat C_{ii\,\cancel{Y_i}}^{d\,LR}}{v_d
  \left( {1 + \dfrac{\alpha_s}{4\pi}C_F+ \tan\beta \hat\epsilon_i^d } \right)}\,,
\label{mb_Yb_NLO}
\end{equation}
with $ \hat C_{fi}^{q\,LR}$ defined in \eq{SigmaHut} and the corresponding equation for  $\hat\epsilon_i^d$. Here $\epsilon_i^{d(1)}$ and  $\epsilon_i^{d\,(2)}$ are defined in direct analogy to  \eq{C1C2}. Further, the Wilson coefficients appearing here are assumed to be in the decoupling limit. Eq.~(\ref{mb_Yb_NLO}) constitutes the NLO determination of the Yukawa coupling of the superpotential. When later inserting the Yukawa coupling into the Wilson coefficients, one has to use this relation\footnote{The generalization to the CKM matrix can be achieved following the procedure of \cite{Crivellin:2008mq,Crivellin:2009sd,Hofer:2009xb}}. 
\medskip

The electroweak contributions (involving charginos and neutralinos) to the relation between the quark masses and the Yukawa couplings are in most regions of the parameter space subleading compared to the strong contributions. However, the LO electroweak corrections are easily as large as the NLO SQCD corrections and should be included in a numerical analysis. This can be achieved by simply adding the corresponding contributions to $\hat C_{ii\,\cancel{Y_i}}^{d\,LR}$ and $\hat\epsilon_i^d$ in \eq{mb_Yb_NLO}.

\section{Effective Higgs vertices}
\label{EFT}

To derive the effective Higgs-quark-quark couplings\footnote{In principle also the renormalization of the 
Higgs potential should be addressed. Our derivation of chirally enhanced flavor effects does not 
depend on the specific relations between Higgs self-couplings and their masses.
Since no chirally enhanced effects occur in the Higgs sector, it is 
consistent to use the tree-level values for the Higgs parameters. 
However, one can as well use the NLO values for the Higgs masses and 
mixing angles which might be even better from the numerical point of view.} we have to assume that the external momenta (flowing through the Higgs-quark-quark vertex) are much smaller than the masses of the virtual SUSY particles running in the loop. This assumption limits the applicability of the resulting Feynman rules. If $m_{H^0},m_{A^0},m_{H^{\pm}}\ll \Msusy$ ($H^0,A^0$, and $H^{\pm}$ denote the neutral CP-even, CP-odd and the charged Higgs boson, respectively),
the effective Feynman rules can be used for the calculation of all flavor-observables (also if the Higgs is propagating in a loop) and for processes with a Higgs on the mass shell. If the hierarchy $m_{H^0},m_{A^0},m_{H^{\pm}}\ll \Msusy$ is not satisfied the effective Higgs vertices can still be used for processes in which the momentum flow through the Higgs-quark-quark vertex is small compared to $\Msusy$ which is true for all low-energy flavor observables with tree-level Higgs exchange (like $B_{d,s}\to\mu^+\mu^-$, $B^+\to\tau^+\nu$ or the double Higgs penguin contributing to $\Delta F=2$ processes).
\medskip

As discussed in the Introduction we use an effective field theory approach in our study of the Higgs-quark-quark couplings which simplifies the calculations significantly. This means that we match the MSSM on the 2HDM of type III at the scale $M_{\rm SUSY}$ rather than calculating the Higgs-quark-quark coupling within the MSSM.
\medskip

Let as first consider the effective Lagrangian of a general 2HDM (including Higgs-quark-quark couplings and kinetic terms):
\begin{equation}
\renewcommand{\arraystretch}{2}
\begin{array}{l}
 {\cal L}^{eff}  = \bar Q_{f\;L}^a \left( {\left( {Y_{fi\,{\rm ew}}^{d\,{\rm tree}}+E_{fi}^{d\;{\rm ew}}  } \right)\epsilon _{ba} H_d^{b\star}  - E_{fi}^{\prime d\;{\rm ew}} H_u^a } \right)d_{i\;R} \\ 
 \phantom{ L^{eff}  =}+ \bar  Q_{f\;L}^a \left( {\left( {Y_{fi\,{\rm ew}}^{u\,{\rm tree}}  + E_{fi}^{u\;{\rm ew}} } \right)\epsilon _{ab} H_u^{b\star}  - E_{fi}^{\prime u\;{\rm ew}} H_d^a } \right)u_{i\;R}  \\ 
  \phantom{ L^{eff}  =}  + \bar d_{f\;R} i\cancel{\partial}\left( {\delta _{fi}  - R_{fi}^{d\,{\rm ew}} } \right)d_{i\;R}  
  + \bar u_{f\;R} i\cancel{\partial}\left( {\delta _{fi}  - R_{fi}^{u\,{\rm ew}} } \right)u_{i\;R}  \\ 
  \phantom{ L^{eff}  =}+ \bar Q_{f\;L}^a i\cancel{\partial}\left( {\delta _{fi}  - L_{fi}^{q\,{\rm ew}} } \right)Q_{i\;L}^a\,,
   \end{array}
 \label{Leff}
\end{equation}
where adding the Hermitian conjugate of the terms involving Higgs
fields is implicitly meant.
The Higgs doublets are defined as
\begin{equation}
\begin{array}{l}
{H_d} = \left( {\begin{array}{*{20}{c}}
{H_d^1}\\
{H_d^2}
\end{array}} \right) = \left( {\begin{array}{*{20}{c}}
{H_d^0}\\
{H_d^ - }
\end{array}} \right)\,,\\
{H_u} = \left( {\begin{array}{*{20}{c}}
{H_u^1}\\
{H_u^2}
\end{array}} \right) = \left( {\begin{array}{*{20}{c}}
{H_u^ + }\\
{H_u^0}
\end{array}} \right)\,.
\end{array}
\end{equation}
In \eq{Leff} $a$, $b$ denote $SU(2)_L$\,-\,indices and $\epsilon_{ab}$ is the
two-dimensional antisymmetric tensor with $\epsilon_{12}=-1$.  We
introduced the holomorphic couplings $E^{q\,{\rm ew}}_{fi}$, the
nonholomorphic couplings $E^{\prime q\,{\rm ew}}_{fi}$ ($q=u,d$), and
the contributions to the kinetic terms $R_{fi}^{d,u\,{\rm ew}}$ and
$L_{fi}^{q\,{\rm ew}}$. Here the superscript ``ew'' refers to the fact
that these terms are given in a weak-interaction eigenbasis. In
\eq{Leff} we already anticipated the MSSM where the terms $E^{(\prime)
  q\,{\rm ew}}_{fi}$, $L_{fi}^{q\,{\rm ew}}$ and $R_{fi}^{q\,{\rm
    ew}}$ are loop induced but $Y^{u\,{\rm tree}}_{fi\,{\rm ew}}$ and
$Y^{d\,{\rm tree}}_{fi\,{\rm ew}}$ are generated at tree level via the
MSSM Yukawa couplings\footnote{In principle, without knowing anything
  about the MSSM, the holomorphic corrections could be absorbed into
  an effective Yukawa coupling (and also the corrections to the
  kinetic terms $R_{fi}^{d,u\,{\rm ew}}$ and $L_{fi}^{q\,{\rm ew}}$
  would not be physical). However, once we go back to the MSSM with
  the SUSY breaking terms as input parameters, also the holomorphic
  corrections become physical.}.

\medskip

To connect the effective theory to the MSSM we go to the super-CKM basis, in which the Yukawa couplings are diagonal, by rotating the fields
\begin{equation}
q_{j\;L,R}  \to U_{ji}^{q\;L,R\left( 0 \right)} q_{i\;L,R}\,,
\end{equation}
such that
\begin{equation}
U_{kf}^{q\;L\left( 0 \right)\star} Y_{kj \,{\rm ew}}^{q\,{\rm tree}} U_{ji}^{q\;R\left( 0 \right)}  = Y^{q_i}_{\rm tree} \delta _{fi} \,.
\end{equation}
We now break the electroweak symmetry and write the effective Lagrangian in component form: 
\begin{equation}
\renewcommand{\arraystretch}{2}
\begin{array}{l}
 {\cal L}^{eff}  = \bar u_{f\;L}^{} V_{fj}^{\left( 0 \right)} \left( {\left( {Y_{\rm tree}^{d_j } \delta _{ji}  + E_{ji}^d } \right)H_d^{2\star}  - E_{ji}^{\prime d} H_u^1} \right)d_{i\;R}  \\ 
 \phantom{ L^{eff}  =} + \bar d_{f\;L}^{} V_{jf}^{\left( 0 \right)\star} \left( {\left( {Y_{\rm tree}^{u_j } \delta _{ji}  + E_{ji}^u } \right)H_u^{1\star}  - E_{ji}^{\prime u} H_d^2} \right)u_{i\;R}  \\ 
\phantom{ L^{eff}  =}  - \bar d_{f\;L}^{} \left( {\left( {Y_{\rm tree}^{d_f } \delta _{fi}  + E_{fi}^d } \right)H_d^{1\star}  + E_{fi}^{\prime d} H_u^2} \right)d_{i\;R}  \\ 
 \phantom{ L^{eff}  =} - \bar u_{f\;L} \left( {\left( {Y_{\rm tree}^{u_f } \delta _{fi}  + E_{fi}^u } \right)H_u^{2\star}  + E_{fi}^{\prime u} H_d^1} \right)u_{i\;R} \\
   \phantom{ L^{eff}  =}  + \bar d_{f\;R} i\cancel{\partial}\left( {\delta _{fi}  - R_{fi}^{d} } \right)d_{i\;R}  
  + \bar u_{f\;R} i\cancel{\partial}\left( {\delta _{fi}  - R_{fi}^{u} } \right)u_{i\;R}  \\ 
  \phantom{ L^{eff}  =}+ \bar d_{f\;L} i\cancel{\partial}\left( {\delta _{fi}  - L_{fi}^{d} } \right)d_{i\;L} 
  + \bar u_{f\;L} i\cancel{\partial}\left( {\delta _{fi}  - L_{fi}^{u}
  } \right)u_{i\;L}\\ 
\phantom{ L^{eff}  =}  - \bar d_{f\;L}^{} \left( {\left( {Y_{\rm tree}^{d_f } \delta _{fi}  + E_{fi}^d } \right)v_d  + E_{fi}^{\prime d} v_u} \right)d_{i\;R}  \\ 
 \phantom{ L^{eff}  =} - \bar u_{f\;L} \left( {\left( {Y_{\rm
       tree}^{u_f } \delta _{fi}  + E_{fi}^u } \right)v_u  +
   E_{fi}^{\prime u} v_d}\right)u_{i\;R} \, , \\  
 \end{array}
 \label{Lcomponents}
\end{equation}
where $V^{\left( 0 \right)}  = U^{u\;L\left( 0 \right)\dag }
U^{d\;L\;\left( 0 \right)} $ is not the physical CKM matrix, but
rather the CKM matrix generated by the misalignment of the Yukawa
couplings. Adding the Hermitian conjugate of the mass terms and the
terms involving Higgs fields is tacitly understood. The terms
\begin{equation}
\renewcommand{\arraystretch}{2}
\begin{array}{l}
 E_{fi}^q  = U_{kf}^{q\;L\left( 0 \right)\star} E_{kj}^{q\,{\rm ew} } U_{ji}^{q\;R\left( 0 \right)}  \\ 
 E_{fi}^{\prime q}  = U_{kf}^{q\;L\left( 0 \right)\star} E_{kj}^{\prime q\, {\rm ew}} U_{ji}^{q\;R\left( 0 \right)}  \\ 
 R_{fi}^{q}  = U_{kf}^{q\;R\left( 0 \right)\star} R_{kj}^{q\, {\rm ew}} U_{ji}^{q\;R\left( 0 \right)}  \\ 
 L_{fi}^d  = U_{kf}^{d\;L\left( 0 \right)\star} L_{kj}^{q\,{\rm ew} } U_{ji}^{d\;L\left( 0 \right)}  \\ 
 L_{fi}^u  = U_{kf}^{u\;L\left( 0 \right)\star} L_{kj}^{q\,{\rm ew} } U_{ji}^{u\;L\left( 0 \right)}= V_{fk}^{\left( 0 \right)} L_{kj}^{d} V_{ij}^{\left( 0 \right)\star} \\ 
 \end{array}
\end{equation}
are now given in the super-CKM basis. Note that this is the same basis as the one in which the effective Lagrangian of \eq{Leff2quark} is given (and the same basis in which we calculated the MSSM contributions to the Wilson coefficients). Thus, comparing the last four lines of \eq{Lcomponents} to \eq{Leff2quark} we have the following relation between the Wilson coefficients and the terms of the 2HDM III Lagrangian (at an arbitrary loop order): 
\begin{equation}
\renewcommand{\arraystretch}{2}
\begin{array}{l}
   E^d_{fi}\,=\,\dfrac{C^{d\,LR}_{fi\,A}}{v_d}\,,\hspace{1.5cm} 
   E^{\prime d}_{fi}\,=\,\dfrac{C^{\prime\, d\,LR}_{fi}}{v_u}\,,\\
   E^u_{fi}\,=\,\dfrac{C^{u\,LR}_{fi\,A}}{v_u}\,,\hspace{1.5cm} 
   E^{\prime u}_{fi}\,=\,\dfrac{C^{\prime\, u\,LR}_{fi}}{v_d}\,,\\
   L^{q}_{fi}\,=C^{q\,LL}_{fi}\,,\hspace{1.5cm} R^{q}_{fi}\,=C^{q\,RR}_{fi}\,.
   \end{array}
   \label{E-Sigma}
\end{equation}
\medskip

Now we want to go to the physical basis with flavor diagonal mass terms and canonical kinetic terms. As a first step we render the kinetic terms canonical by a field redefinition:
\begin{equation}
\begin{array}{l}
 q_{i\;L}  \to \left( {\delta _{ij}  + \dfrac{1}{2}L_{ij}^{q} } \right)q_{j\;L} \,, \\ 
 q_{i\;R}  \to \left( {\delta _{ij}  + \dfrac{1}{2}R_{ij}^{q} } \right)q_{j\;R} \,. \\ 
 \label{kinetic}
 \end{array}
\end{equation}
Consider now the quark mass matrices. The redefinition of the fields in \eq{kinetic} also leads to a shift in down-quark mass matrix so that it is now given by 
\begin{equation}
\renewcommand{\arraystretch}{2}
\begin{array}{l}
m_{fi}^d  = \left( {\hat{\hat E}_{fi}^d  + \hat{\hat Y}_{fi}^{d\,{\rm tree}} } \right)v_d  + v_u \hat{\hat E}_{fi}^{\prime d } = \hat{\hat C} _{fi}^{d\;LR}  + v_d \hat{\hat Y}_{fi}^{d\,{\rm tree}} \,,\\
m_{fi}^u  = \left( {\hat{\hat E}_{fi}^u  + \hat{\hat Y}_{fi}^{u\,{\rm tree}} } \right)v_u  + v_d \hat{\hat E}_{fi}^{\prime u } = \hat{\hat C} _{fi}^{u\;LR}  + v_u \hat{\hat Y}_{fi}^{u\,{\rm tree}} \,,
\end{array}
\end{equation}
where we have defined 
\begin{equation}
\renewcommand{\arraystretch}{2}
\begin{array}{l}
 \hat {\hat E}_{fi}^{(\prime)q}  = E_{fi}^{(\prime)q}  +
 \dfrac{1}{2}\sum\limits_{j = 1}^3 {\left( {L_{fj}^{q}
     E_{ji}^{(\prime)q}  + E_{fj}^{(\prime)q} R_{ji}^{q} } \right)}
 \,, \\ 
 \hat{ \hat Y}_{fi}^{q\,{\rm tree}}  = Y^{q_i}_{{\rm tree}}\delta_{fi}
 + \dfrac{1}{2}\sum\limits_{j = 1}^3 {\left( {C_{fj}^{q\,LL}
     Y^{q_i}_{\rm tree}\delta_{ji}  + Y^{q_f}_{\rm tree}\delta_{fj}
     C_{ji}^{q\,RR} } \right)}  \,, \\ 
 \hat{ \hat C} _{fi}^{q\;LR}  = C _{fi}^{q\;LR}  +
 \dfrac{1}{2}\sum\limits_{j = 1}^3 {\left( {C_{fj}^{q\,LL} C
     _{ji}^{q\;LR}  + C _{fj}^{q\;LR} C_{ji}^{q\,RR} } \right)}  \,. \\ 
 \end{array}
\end{equation}
Note that the quantities with a double hat contain also the contributions from flavor-changing LL and RR Wilson coefficients, while the quantities with one hat (see \eq{SigmaHut} and \eq{EHut}) only contain the flavor-conserving LL and RR Wilson coefficients. 
\medskip

We now diagonalize the quark mass matrices by a bi-unitary transformation
\begin{equation}
U_{kf}^{q\,L\star} m_{kj}^q U_{ji}^{q\,R}  = m_{q_i } \delta _{fi} \,,
\label{m_diagonalization}
\end{equation}
where the rotation matrices 
\begin{equation}
\renewcommand{\arraystretch}{2}
U_{}^{q\,L}  = \left( {\begin{array}{*{20}c}
   1 & {\dfrac{{m_{12}^q }}{{m_{q_2 } }}} & {\dfrac{{m_{13}^q }}{{m_{q_3 } }}}  \\
   {\dfrac{{ - m_{12}^{q\star} }}{{m_{q_2 } }}} & 1 & {\dfrac{{m_{23}^q }}{{m_{q_3 } }}}  \\
   {\dfrac{{ - m_{13}^{q\star} }}{{m_{q_3 } }} + \dfrac{{m_{12}^{q\star} m_{23}^{q\star} }}{{m_{q_2 } m_{q_3 } }}} & {\dfrac{{ - m_{23}^{q\star} }}{{m_{q_3 } }}} & 1  \\
\end{array}} \right),\;\;\;U_{}^{q\,R}  = \left( {\begin{array}{*{20}c}
   1 & {\dfrac{{m_{21}^{q\star} }}{{m_{q_2 } }}} & {\dfrac{{m_{31}^{q\star} }}{{m_{q_3 } }}}  \\
   {\dfrac{{ - m_{21}^q }}{{m_{q_2 } }}} & 1 & {\dfrac{{m_{32}^{q\star} }}{{m_{q_3 } }}}  \\
   {\dfrac{{ - m_{31}^q }}{{m_{q_3 } }} + \dfrac{{m_{32}^q m_{21}^q }}{{m_{q_2 } m_{q_3 } }}} & {\dfrac{{ - m_{32}^{q\star} }}{{m_{q_3 } }}} & 1  \\
\end{array}} \right)
\label{DeltaU}
\end{equation}
are obtained from a perturbative diagonalization of the quark mass matrix\footnote{Note that these rotations are identical
  to the ones obtained in the diagrammatic approach (see Ref.~\cite{Crivellin:2010er} for details).}.
\medskip

Switching to the physical basis in which the quark mass matrices are diagonal, these rotations modify the effective Lagrangian as follows~\cite{Crivellin:2010er}:
\bea
\renewcommand{\arraystretch}{2.2}
\begin{array}{l}
\mathcal{L}^{eff}  = \bar u_{f\;L}^{} U_{kf}^{u\;L\star} V_{kk^\prime}^{\left( 0 \right)} \left( {\dfrac{{m_{k^\prime j}^d }}{{v_d }}H_d^{2\star}  - \hat{\hat E}_{k'j}^{\prime d} \left( {H_u^1  + \tan \left( \beta  \right)H_d^{2\star} } \right)} \right)U_{ji}^{d\;R} d_{i\;R}  \\ 
\phantom{\mathcal{L}^{eff}  =}  + \bar d_{f\;L}^{} U_{kf}^{d\;L\star} V_{k^\prime k}^{\left( 0 \right)\star} \left( {\dfrac{{m_{k^\prime j}^u }}{{v_u }}H_u^{1\star}  - \hat{\hat E}_{k'j}^{\prime u} \left( {H_d^2  + \cot \left( \beta  \right)H_u^{1\star} } \right)} \right)U_{ji}^{d\;R} u_{i\;R}  \\ 
\phantom{\mathcal{L}^{eff}  =}  - \bar d_{f\;L}^{} U_{kf}^{d\;L\star} \left( {\dfrac{{m_{kj}^d }}{{v_d }}H_d^{1\star}  + \hat{\hat E}_{kj}^{\prime d} \left( {H_u^2  - \tan \left( \beta  \right)H_d^{1\star} } \right)} \right)U_{ji}^{d\;R} d_{i\;R}  \\ 
\phantom{\mathcal{L}^{eff}  =}  - \bar u_{f\;L}^a U_{kf}^{u\;L\star} \left( {\dfrac{{m_{kj}^u }}{{v_u }}H_u^{2\star}  + \hat{\hat E}_{kj}^{\prime u} \left( {H_d^1  - \cot \left( \beta  \right)H_u^{2\star} } \right)} \right)U_{ji}^{u\;R}u_{i\;R} \,+\,h.c.  \\ 
 \end{array}
\label{L-Y-FCNC}
\eea
where we skipped the mass terms and the kinetic terms. This can be further simplified by using the physical CKM matrix given by
\begin{equation}
V_{fi}=	U_{jf}^{u\;L\star} V_{jk}^{\left( 0 \right)} U_{k i}^{d\;L}\,.
\label{CKMphys}
\end{equation}
In addition, we define the abbreviations 
\begin{eqnarray}
\renewcommand{\arraystretch}{2.0}
\begin{array}{l}
 \tilde E_{fi}^{\prime q}  = U_{kf}^{q\;L\star} \hat {\hat E}_{kj}^{\prime q} U_{ji}^{q\;R}  \\ 
\phantom{ \tilde E_{fi}^{\prime q} }  = \hat E_{fi}^{\prime q}  - \left( {\begin{array}{*{20}c}
   0 & {\hat E_{22}^{\prime q} \hat c _{12}^{q\;LR} } & {\hat E_{33}^{\prime q} \left( {\hat c _{13}^{q\;LR}  - \hat c _{12}^{q\;LR} \hat c _{23}^{q\;LR} } \right)}  \\
   {\hat E_{22}^{\prime q} \hat c _{21}^{q\;LR} } & 0 & {\hat E_{33}^{\prime q} \hat c _{23}^{q\;LR} }  \\
   {\hat E_{33}^{\prime q} \left( {\hat c _{31}^{q\;LR}  - \hat c _{32}^{q\;LR} \hat c _{31}^{q\;LR} } \right)} & {\hat E_{33}^{\prime q} \hat c _{32}^{q\;LR} } & 0  \\
\end{array}} \right)_{fi} \\
 \phantom{ \tilde E_{fi}^{\prime q} }  \equiv \hat E_{fi}^{\prime q}  - \Delta \hat E_{fi}^{\prime q} \,. 
 \end{array}
\label{Etilde}
\end{eqnarray}
Note that in this expression only quantities with a single hat defined as
\begin{equation}
 \hat E_{fi}^{(\prime)q}  = E_{fi}^{(\prime)q}  + \frac{1}{2}\left( {L_{ff}^{q} E_{fi}^{(\prime)q}  + E_{fi}^{(\prime)q} R _{ii}^{q} } \right)\,´,
 \label{EHut}
\end{equation}
and $\hat c^{q\;LR}_{ij}$ defined by combining \eq{SigmaHut} with 
\begin{equation}
   c^{q\;LR}_{ji}\,=\,\dfrac{C_{ji}^{q\,LR}}{\max\{m_{q_j},m_{q_i}\}}\,,
\label{eq:sigdef}
\end{equation}
enter. This is in agreement with the finding of Ref.~\cite{Gorbahn:2009pp} that the effect of the flavor-changing LL and RR self-energies drops out in the effective Higgs vertices.
\medskip

Finally, to arrive at the effective Feynman rules we project the fields $H^0_u$ and $H^0_d$ onto the physical components
$H^0$, $h^0$, $A^0$ and $H^\pm$ as
\begin{eqnarray}
H_u^0&=&\frac{1}{\sqrt{2}}\left(H^0\sin\alpha + h^0\cos\alpha +
iA^0\cos\beta\right)\,, \nonumber\\
H_d^0&=&\frac{1}{\sqrt{2}}\left(H^0\cos\alpha - h^0\sin\alpha +
iA^0\sin\beta\right)\,,\nonumber\\
H_u^{1\star} &=& \cos \left( \beta  \right){H^ - }\,,\nonumber\\
H_d^2 &=& \sin \left( \beta  \right){H^ - }\,.
  \label{Htilde}
\end{eqnarray}\medskip
\medskip

Using \eq{CKMphys}, \eq{Etilde}, and \eq{Htilde}, the effective Lagrangian in
\eq{L-Y-FCNC} leads to the following effective Higgs-quark-quark
Feynman rules\footnote{Note that some of the Higgs-quark-quark
  couplings are suppressed by a factor $\cos\beta$ or $\sin\alpha$
  stemming from the Higgs mixing matrices. If one decides to keep
  these suppressed couplings, one should be aware of the fact that
  they receive proper vertex corrections in which the suppression
  factor does not occur and which are thus $\tan\beta$ enhanced with
  respect to the tree-level couplings. Such enhanced corrections to
  the coupling of $H^{\pm}$ to right-handed up quarks are important
  for $b\to s \gamma$ \cite{Carena:2000uj,Degrassi:2000qf}.} shown in Fig.~\ref{fig:Higgs-Quark-Coupling} (note
that the CKM matrix $V$ in the charged Higgs coupling
is the physical one):
\begin{eqnarray}
{\Gamma_{u_f u_i }^{LR\,H_k^0} } &=& x_u^k\left( \frac{m_{u_i }}{v_u}
\delta_{fi} - \widetilde E_{fi}^{\prime u}\cot\beta \right) + x_d^{k\star}
\widetilde E_{fi}^{\prime u}\,, \nonumber\\[0.2cm]
{\Gamma_{d_f d_i }^{LR\,H_k^0} } &=& x_d^k \left( \frac{m_{d_i
}}{v_d} \delta_{fi} - \widetilde E_{fi}^{\prime d}\tan\beta \right) +
x_u^{k\star}\widetilde E_{fi}^{\prime d} \,,\nonumber \\[0.2cm]
{\Gamma_{u_f d_i }^{LR\,H^\pm} } &=& \sum\limits_{j = 1}^3
{\sin\beta\, V_{fj} \left( \frac{m_{d_i }}{v_d} \delta_{ji}-
  \widetilde{E}^{\prime d}_{ji}\tan\beta \right)\,, }
\nonumber\\[0.2cm]
{\Gamma_{d_f u_i }^{LR\,H^ \pm } } &=& \sum\limits_{j = 1}^3
{\cos\beta\, V_{jf}^{\star} \left( \frac{m_{u_i }}{v_u} \delta_{ji}-
  \widetilde{E}^{\prime u}_{ji}\tan\beta \right)\, }\,,
 \label{Higgs-vertices-decoupling}
\end{eqnarray}
where for $H^0_k=(H^0,h^0,A^0)$ the coefficients $x_q^{k}$ are given
by
\begin{equation}
x_d^k \, = \,\left(-\frac{1}{\sqrt{2}}\cos\alpha,\,\frac{1}{\sqrt{2}}\sin\alpha,
\,\frac{i}{\sqrt{2}}\sin\beta\right), \hspace{1cm}
x_u^k \, = \, \left(-\frac{1}{\sqrt{2}}\sin\alpha,\,-\frac{1}{\sqrt{2}}\cos\alpha,
\,\frac{i}{\sqrt{2}}\cos\beta\right)\,.
\end{equation}
It is important to keep in mind that the $\hat c^{q\;LR}_{ij}$ in
\eq{Etilde} must be calculated using the quantities $Y^{q}$
and $V^{(0)}$ of the MSSM superpotential.
\medskip

\begin{nfigure}{t}
\includegraphics[width=0.9\textwidth]{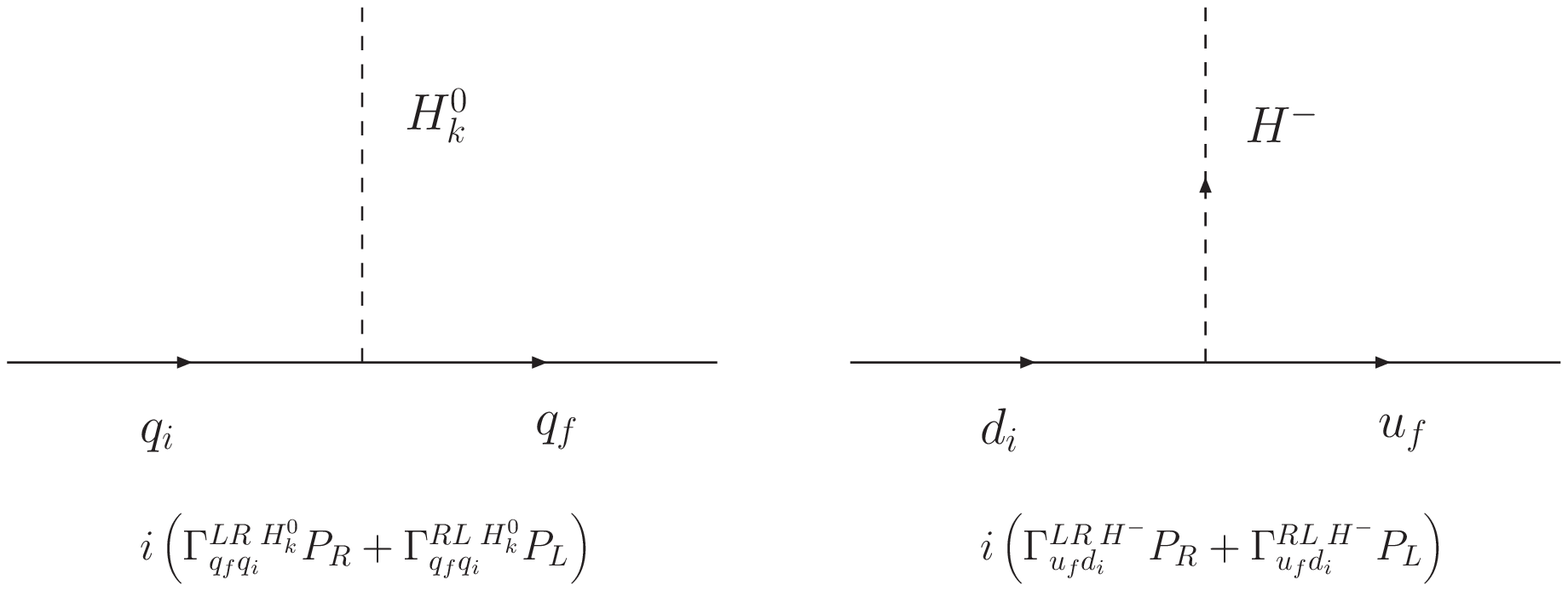}
\caption{Higgs-quark vertices with the corresponding Feynman rules. The couplings with exchanged chirality structure are obtained from \eq{Higgs-vertices-decoupling} by using ${\Gamma_{q_f q_i }^{RL\,H} }=\Gamma_{q_i q_f }^{LR\,H\star} $.  }
\label{fig:Higgs-Quark-Coupling}
\end{nfigure}

Note that without the nonholomorphic corrections $E^{\prime q}_{ij}$ the
rotation matrices $U^{q\,L,R}$ would simultaneously diagonalize the
effective mass terms and the neutral Higgs couplings in \eq{L-Y-FCNC}.
However, in the presence of nonholomorphic corrections this is no
longer the case and apart from a flavor-changing nonholomorphic
correction a term proportional to a flavor-conserving
nonholomorphic correction times a flavor-changing self-energy is also
generated.
\medskip

\subsection{Effective Higgs-quark-quark vertices at NLO}

The effective Higgs-quark-quark vertices at NLO in the MSSM are obtain in the following way: After inserting the definition for $\tilde E^{\prime q}_{fi}$ (see \eq{Etilde}) into \eq{Higgs-vertices-decoupling} we express $\hat E^{(\prime) q}_{fi}$ through $\hat C^{(\prime) q}_{fi}$ according to \eq{E-Sigma}.

\section{Conclusions}
\label{conclusions}
In this article we computed the genuine two-loop SQCD corrections to
the chirality-changing quark self-energies.  In the limit where the
external momentum and the quark mass are zero, we
presented relatively simple analytic results without making further
assumptions on the SUSY spectrum. Because of the one-to-one correspondence
(in the decoupling limit) between chirality-changing quark
self-energies and Higgs-quark-quark vertices, this is an efficient and
elegant way of calculating at the same time not only effective Higgs
vertices, but also the Yukawa couplings and CKM elements of the MSSM
superpotential in terms of the physical quark masses and the physical 
CKM matrix.

\medskip

Our next-to-leading order results increase the values of Wilson coefficients $C^{q\,LR}_{fi}$ of the operators $\overline{q}_f \, P_{R} \, q_i$ by approximately $9\%$ compared to the values obtained at leading order. This means that, since at large $\tan\beta$ the threshold corrections to the Yukawa couplings of the two-loop correction is ${\cal O}(10\%)$.  At the same time the matching scale
uncertainty of the effective Higgs-quark-quark couplings and of the
corresponding Wilson coefficients is significantly reduced (see Fig.~\ref{mu-abhaengigkeit}). 

\medskip

We resummed all chirally enhanced corrections modifying the relation
between the quark masses and the Yukawa couplings of the MSSM
superpotential up to order $\alpha_s^{n+1}\tan^n\beta$ (see \eq{mb_Yb_NLO}). The resulting MSSM Yukawa couplings can be used for a precision study of Yukawa
unification. Furthermore, using these Yukawa couplings, we derived
effective Higgs-quark-quark vertices (see \eq{Higgs-vertices-decoupling}) entering the calculation of FCNC processes and also of Higgs decays, as long as the momentum transfer is
small compared to the SUSY scale.
\medskip

\begin{acknowledgments}
This work is supported by the
Swiss National Science Foundation. 
A. C. thanks Ulrich Nierste for help in the early stages
of this project. We thank Youichi Yamada for checking our results and finding a typesetting mistake in Eq.~(25) in the previous version of this article.
\end{acknowledgments}

\begin{appendix}

\section{One-loop results}\label{sec:LOresult}
Here we summarize various one-loop results necessary for the two-loop
calculation of the chirality flipping self-energy (see
\cite{Greub:2011ji} for details). 
Unless stated otherwise, all
expressions appearing in this appendix were obtained in dimensional 
regularization. The matrices $W^q$ diagonalize the squark mass matrices according to \eq{Wdef} and we use the definitions:
\begin{equation}
	x_s=\dfrac{m_{\tilde d_s}}{m_{\tilde g}}\,, \qquad y_s=\dfrac{m_{\tilde u_s}}{m_{\tilde g}}\,, \qquad x_\mu=\dfrac{\mu}{m_{\tilde g}}\,.
\end{equation}

\subsection{Self-energies}
Here we give the explicit one-loop results for quark, gluino, and
squark self-energies in dimensional regularization, where we put
$D=4-2\varepsilon$ and write the renormalization scale in the form
$\mu e^{\gamma/2}/(\sqrt{4\pi})$.
Our conventions are such that the calculation of the truncated
self-energy diagrams give $-i\Sigma$.

\subsubsection{Quark}
The one-loop quark self-energies induced by gluinos and squarks are given by
\begin{equation}
\renewcommand{\arraystretch}{2.4}
\begin{array}{l}
  \Sigma_{q_f q_i }^{\tilde g\;LR} (0)   =
  \dfrac{\alpha_s}{2\pi}
 W_{fs}^{\tilde q} W_{i + 3,s}^{\tilde q\star} \,
C_F \, m_{\tilde g} \, B_0(0;m_{\tilde{g}}^2,m_{\tilde{q}_s}^2) \\
\phantom{\Sigma_{q_f q_i }^{\tilde g\;LR} (0)}  = \dfrac{\alpha_s}{2\pi}
 W_{fs}^{\tilde q} W_{i + 3,s}^{\tilde q\star} \,
C_F \, m_{\tilde g}
\left( {\dfrac{{x_s^2 \ln \left( {x_s^2 } \right)}}{{1 - x_s^2 }} }\right.\\
 \qquad\qquad\qquad\qquad\qquad \left. {- \varepsilon \dfrac{{x_s^2
       \left( {\ln ^2 \left( {x_s^2 } \right) - 2\ln \left( {x_s^2 }
         \right) - 2\ln \left( {x_s^2 } \right)\ln \left( {x_\mu ^2 }
         \right)} \right)}}{{2\left( {1 - x_s^2} \right)}} + O\left( {\varepsilon ^2 } \right)} \right)\,, \\ 

 \Sigma_{q_f q_i }^{\tilde g\;LL} (0)  =
   \dfrac{\alpha_s}{2\pi}
 W_{fs}^{\tilde q} W_{i,s}^{\tilde q\star} \,
C_F \, B_1(0;m_{\tilde{g}}^2,m_{\tilde{q}_s}^2) \\
\phantom{\Sigma _{q_f q_i }^{\tilde g\;LL}  (0)}   = - \dfrac{{ \alpha _s
 }}{{4\pi }}  W_{fs}^{\tilde q} W_{is}^{\tilde q\star} C_F \left( {\dfrac{1}{\varepsilon }  + \ln \left( {x_\mu ^2 } \right) + \dfrac{{3 - 4x_s^2  + x_s^4  + \left( {4x_s^2  - 2x_s^4 } \right)\ln \left( {x_s^2 } \right)}}{{2\left( {1 - x_s^2 } \right)^2 }}} \right) 
 \\ \qquad\qquad\;\;\;\; +\; O\left( \varepsilon  \right)\,. \\ 
 \end{array}
 \label{SigmaLR1}
\end{equation}
Using unitarity, we can replace 
$ B_0(0;m_{\tilde{g}}^2,m_{\tilde{q}_s}^2) $
by $ [B_0(0;m_{\tilde{g}}^2,m_{\tilde{q}_s}^2) -
B_0(0;m_{\tilde{g}}^2,0)]$
in the first line of  $\Sigma_{q_f q_i }^{\tilde g\;LR} (0)$. 
This we did when writing the explicit expression.

\noindent
The ordinary gluon correction reads in Feynman gauge
\begin{equation}
\renewcommand{\arraystretch}{2}
\begin{array}{l}
 \Sigma _{q_f q_i }^{g\;LL,RR} \left( p^2 \right) =
 \dfrac{\alpha_s}{{4\pi }} {C_F}\left( {d-2} \right){B_1}\left(
       {{p^2};m_{{q_i}}^2,0} \right) \, \delta_{fi} \,, \\ 
  \phantom{\Sigma _{q_f q_i }^{g\;LL,RR}\left( p^2 \right)} =
  \dfrac{\alpha_s}{{4\pi }} {C_F}\left( { - \dfrac{1}{\varepsilon } +
    \dfrac{\left(p^2\right)^2 - m_{{q_i}}^4}{\left(p^2\right)^2}
    \ln \left( \dfrac{m_{{q_i}}^2 - {p^2}-i0}{m_{{q_i}}^2} \right) -
\dfrac{{m_{{q_i}}^2}}{{{p^2}}} + \ln \left(
      {\dfrac{{m_{{q_i}}^2}}{{{\mu ^2}}}} \right) - 1} \right) \,
  \delta_{fi} \\
  \phantom{\Sigma _{q_f q_i }^{g\;LL,RR}\left( p^2 \right) =}  + O(\varepsilon)\, , \\ 
 \end{array}	\label{quark_gluon_SE1}
\end{equation}
\begin{equation}
\renewcommand{\arraystretch}{2}
\begin{array}{l}
 \Sigma _{q_f q_i }^{g\;LR,RL} \left( p^2 \right) =
 \dfrac{\alpha_s}{4\pi} \, C_F \, d \, {m_{{q_i}}} \, {B_0}\left(
       {{p^2};m_{{q_i}}^2,0} \right)\, \delta_{fi} \, , \\ 
\phantom{ \Sigma _{q_f q_i }^{g\;LR,RL}  \left( p^2 \right)}  =
\dfrac{\alpha_s}{\pi } \, C_F \, m_{q_i} \left(
      \dfrac{1}{\varepsilon } + \dfrac{m_{q_i}^2 - p^2}{p^2} \ln \left(
            \dfrac{m_{q_i}^2 - p^2 -i0}{m_{q_i}^2} \right) -
        \ln \left( \dfrac{m_{q_i}^2}{\mu ^2} \right) +
        \dfrac{3}{2} \right) \, \delta_{fi}\\
\phantom{ \Sigma _{q_f q_i }^{g\;LR,RL}  \left( p^2 \right)  = }        + O(\varepsilon) \, . \\ 
 \end{array}	\label{quark_gluon_SE2}
\end{equation}
Note that \eq{quark_gluon_SE1} and \eq{quark_gluon_SE2} are given in
dimensional regularization (not dimensional reduction).

\subsubsection{Gluino}
Here we assume that of the three gaugino masses the gluino mass is chosen to be real which is always possible. For the gluino self-energy the part induced by a gluon reads
\begin{equation}
\Sigma _{\tilde g\tilde g}^g \left( p^2 \right) = \dfrac{{\alpha _s
}}{{4\pi }}C_A \left( d\,m_{\tilde g} B_0 \left( p^2;m_{\tilde g}^2,0
  \right) + \cancel{p}\left( {d-2} \right) B_1 \left( {p^2;m_{\tilde
      g}^2,0 } \right) \right) \, ,
\end{equation}
which decomposes for on-shell gluinos into
\begin{equation}
\renewcommand{\arraystretch}{2}
\begin{array}{l}
 \Sigma _{\tilde g\tilde g}^{g\;LR,RL} \left( {m_{\tilde g}^2 } \right) = \dfrac{{\alpha _s }}{\pi }C_A \,m_{\tilde g} \left( {\dfrac{1}{\varepsilon } + \dfrac{3}{2} + \ln \left( {x_\mu ^2 } \right)} \right) + O\left( \varepsilon  \right)\,, \\ 
 \Sigma _{\tilde g\tilde g}^{g\;LL,RR} \left( {m_{\tilde g}^2 }
 \right) = -\dfrac{{ \alpha _s }}{{4\pi }}C_A \left(
       {\dfrac{1}{\varepsilon }
 + 2 + \ln \left( {x_\mu ^2 } \right)} \right) + O\left( \varepsilon  \right)\,, \\ 
 \end{array}
\end{equation}
where we inserted the explicit expressions for the loop functions.
The part of the gluino self-energy with squarks and quarks as virtual particles in the approximation $m_q=0$ is given by
\begin{eqnarray}
\Sigma _{\tilde g\tilde g}^{q\tilde q\,LR,RL} \left( p^2 \right) &=& 0
\, ,
\nonumber \\
\Sigma _{\tilde g\tilde g}^{q\tilde q\,LL,RR} \left( p^2 \right) &=&
\dfrac{{\alpha_s}}{{4\pi }} \, 2 \, tr \, \sum\limits_{s = 1}^6 {\left( {B_1 \left( {p^2 ;0,m_{\tilde d_s }^2 } \right) + B_1 \left( {p^2 ;0,m_{\tilde u_s }^2 } \right)} \right)}  \,,
\end{eqnarray}
where the latter reads explicitly for on-shell gluinos
\begin{equation}
\renewcommand{\arraystretch}{2}
\begin{array}{l}
\Sigma _{\tilde g\tilde g}^{q\tilde q\,LL,RR} \left( m_{\tilde{g}}^2 \right) = -
\dfrac{{\alpha_s}}{{4\pi }} \, 2 \, tr \, \left[ {\left( {\dfrac{1}{\varepsilon } + 2 + \ln \left( {x_\mu ^2} \right)} \right){n_f} }\right.\\
 \phantom{\Sigma _{\tilde g\tilde g}^{q\tilde q\,LL,RR} \left( m_{\tilde{g}}^2 \right) =} \left.{- \dfrac{1}{2}\sum\limits_{s = 1}^6\left(
  {x_s^2 + {{\left( {1 - x_s^2} \right)}^2}\ln \left( {\dfrac{{x_s^2 - 1 -i\,0}}{{x_s^2}}} \right) + \ln \left( {x_s^2} \right) \;+\; (x_s \to y_s)} \right)} \right]\\
 \phantom{\Sigma _{\tilde g\tilde g}^{q\tilde q\,LL,RR} \left( m_{\tilde{g}}^2 \right) =}  + O\left( \varepsilon  \right)\,,
\end{array}
\end{equation}
with $n_f=6$. The quantities $\Sigma_{\tilde g}^{LL}(m_{\tilde{g}}^2)$ and
$\Sigma_{\tilde g}^{LR}(m_{\tilde{g}}^2)$ that appear in
eq. (\ref{gluinomassrenorm}) are defined as
\begin{equation}
\Sigma_{\tilde g}^{LL,LR}(m_{\tilde{g}}^2)=
\Sigma _{\tilde g\tilde g}^{g\;LL,LR} \left( {m_{\tilde g}^2 } \right) +
\Sigma _{\tilde g\tilde g}^{q\tilde{q}\;LL,LR} \left( {m_{\tilde g}^2
} \right) \, .
\end{equation}

\subsubsection{Squark}
For the squark self-energy we have
\begin{equation}
\Sigma _{\tilde q_s \tilde q_t } =\Sigma _{\tilde q_s \tilde q_t }^g+\Sigma _{\tilde q_s \tilde q_t }^{\tilde gq} +\Sigma _{\tilde q_s \tilde q_t }^{\tilde q}\,,
\label{squark_sum}
\end{equation}
where the parts refer to the squark self-energy with gluon
\begin{eqnarray}
\Sigma _{\tilde q_s \tilde q_t }^g \left( {p^2 } \right) = \dfrac{\alpha_s}{{4\pi }} C_F \left( {2\left( {p^2  + m_{\tilde q_s }^2 } \right)B_0 \left( {p^2 ;m_{\tilde q_s }^2,0 } \right) - A_0 \left( {m_{\tilde q_s }^2 } \right)} \right)\delta _{st} \,, \\
\Sigma _{\tilde q_s \tilde q_t }^g \left( {m_{\tilde q_s }^2 } \right)
= 3\, \dfrac{\alpha_s}{{4\pi }} C_F m_{\tilde q_s }^2 \left( {\dfrac{{\rm{1}}}{\varepsilon} - \ln \left( {x_s^2 } \right) + \ln \left( {x_\mu ^2 } \right) + \dfrac{7}{3}} \right)\delta _{st}  + O\left( \varepsilon  \right) \,,
\label{squark_gluon_SE}
\end{eqnarray}
the squark self-energy with quark and gluino
\begin{equation}
\renewcommand{\arraystretch}{2}
\begin{array}{l}
\Sigma _{\tilde q_s \tilde q_t }^{\tilde gq} \left( {p^2 } \right) = \dfrac{\alpha_s}{{2\pi }} C_F \left( {A_0 \left( {m_{\tilde g}^2 } \right) + \left( {m_{\tilde g}^2  - p^2 } \right)B_0 \left( {p^2 ;m_{\tilde g}^2,0 } \right)} \right)\delta _{st} \,,\\
\Sigma _{\tilde q_s \tilde q_t }^{\tilde gq} \left( {m_{\tilde q_s }^2
} \right) = \dfrac{\alpha_s}{{2\pi }} C_F \, m_{\tilde g}^2 \left[
  \dfrac{{2 - x_s^2 }}{\varepsilon } + 3 - 2 \, x_s^2  + \left( {2 - x_s^2} \right)\ln \left( {x_\mu ^2 } \right) \right.\\
\phantom{\Sigma _{\tilde q_s \tilde q_t }^{\tilde gq} \left( {m_{\tilde q_s }^2 } \right) =}\qquad
\left. + \left( {\dfrac{1}{{x_s^2 }} + x_s^2  - 2} \right)\ln \left(
     {1 - x_s^2 -i0 } \right) \right]\delta _{st}  + O\left( \varepsilon  \right) \,,
\end{array}
\label{squark_gluino_SE}
\end{equation}
and the squark tadpole self-energy of Fig.~\ref{squark-tadpole} (for up (down) type squarks only the diagram with internal up (down) squarks is nonzero):
\begin{equation}
\renewcommand{\arraystretch}{2}
\begin{array}{l}
 \Sigma _{\tilde q_s \tilde q_t }^{\tilde q\tilde q}  = -\dfrac{{\alpha_s}}{{4\pi }}C_F \left( {\delta _{st} A_0 \left( {m_{\tilde q_s }^2 } \right) }\right.\\
\phantom{ \Sigma _{\tilde q_s \tilde q_t }^{\tilde q\tilde q}  =} \left.{- 2\sum\limits_{i,j = 1}^3 {\sum\limits_{s' = 1}^6 {\left( {W_{i + 3s}^{\tilde q\star} W_{i + 3s'}^{\tilde q} W_{js'}^{\tilde q\star} W_{jt}^{\tilde q}  + W_{is}^{\tilde q\star} W_{is'}^{\tilde q} W_{j + 3s'}^{\tilde q\star} W_{j + 3t}^{\tilde q} } \right)A_0 \left( {m_{\tilde q_{s'} }^2 } \right)} } } \right) \\ 
 \begin{array}{*{20}c}
   { \phantom{ \Sigma _{\tilde q_s \tilde q_t }^{\tilde q\tilde q}}= - \dfrac{{\alpha_s}}{{4\pi }} C_F \left[ {\delta _{st} m_{\tilde q_s }^2 \left( {\dfrac{1}{\varepsilon } + 1 - \ln \left( {x_s^2 } \right) + \ln \left( {x_\mu ^2 } \right)} \right)} \right.} \hfill  \\
\phantom{ \Sigma _{\tilde q_s \tilde q_t }^{\tilde q\tilde q}  =} - 2\sum\limits_{i,j = 1}^3 \sum\limits_{s' = 1}^6 \left( {W_{i + 3s}^{\tilde q\star} W_{i + 3s'}^{\tilde q} W_{js'}^{\tilde q\star} W_{jt}^{\tilde q}  + W_{is}^{\tilde q\star} W_{is'}^{\tilde q} W_{j + 3s'}^{\tilde q\star} W_{j + 3t}^{\tilde q} } \right)m_{\tilde q_{s'} }^2 ´\\
\left.\qquad\qquad\qquad\qquad\;\;\;\times\left( {\dfrac{1}{\varepsilon } + 1 - \ln \left( {x_{s'}^2 } \right) + \ln \left( {x_\mu ^2 } \right)} \right)    { + O\left( {\varepsilon} \right)} \right] \hfill  \\
\end{array}
\end{array} 
\label{squark_squark_SE}
\end{equation}
Note that $\Sigma _{\tilde q_s \tilde q_t }^{\tilde q\tilde q}$ is
independent of the external momentum.
The part proportional to $\delta_{st}$ in \eq{squark_squark_SE} is due to diagram b) of Fig.~\ref{squark-tadpole}
while the second part, which is proportional to at least one element $\Delta^{q\,LR}_{ij}$, is generated by diagram a). 
\medskip

Note that in the sum of all contributions to the
diagonal squark self-energy there is no
divergence proportional to $p^2$ and thus no wave-function
renormalization is needed in
order to render the diagonal squark two point function finite.

\subsection{Loop functions}
The one-loop functions $A_0(m^2)$, $B_0(p^2;m_1^2,m_2^2)$, and
$B_1(p^2;m_1^2,m_2^2)$ in the previous paragraph are defined as
\begin{equation}
A_0(m^2) = \dfrac{16 \pi^2}{i} \dfrac{\mu^{2\varepsilon}
  e^{\gamma \varepsilon}}{(4\pi)^{\varepsilon}} \, \int \dfrac{d^d
  \ell}{(2\pi)^d} \, \dfrac{1}{[\ell^2-m^2]}
\end{equation}
\begin{equation}
B_0(p^2;m_1^2,m_2^2) = \dfrac{16 \pi^2}{i} \dfrac{\mu^{2\varepsilon}
  e^{\gamma \varepsilon}}{(4\pi)^{\varepsilon}} \, \int \dfrac{d^d
  \ell}{(2\pi)^d} \, \dfrac{1}{[\ell^2-m_1^2] \, [(\ell+p)^2-m_2^2]}
\end{equation}
\begin{equation}
B_1(p^2;m_1^2,m_2^2) \, p^{\mu} = \dfrac{16 \pi^2}{i} \dfrac{\mu^{2\varepsilon}
  e^{\gamma \varepsilon}}{(4\pi)^{\varepsilon}} \, \int \dfrac{d^d
  \ell}{(2\pi)^d} \, \dfrac{\ell^{\mu}}{[\ell^2-m_1^2] \, [(\ell+p)^2-m_2^2]}
\end{equation}
The function $B_0(m_1^2,m_2^2)$ which also appears, is an abbreviation
for $B_0(0;m_1^2,m_2^2)$. We give now relations among these functions
and explicit versions for specific arguments
\begin{equation}
\renewcommand{\arraystretch}{2.4}
\begin{array}{l}
 A_0 \left( {m^2 } \right) = m^2 \left[ {\dfrac{1}{\varepsilon } + 
\ln \left( {\dfrac{{\mu^2 }}{{m^2 }}} \right) + 1}  
+ \left(\dfrac{\pi^2}{12} +1 + \ln \left( {\dfrac{{\mu^2 }}{{m^2 }}}
\right)+
\dfrac{1}{2}\ln^2 \left( {\dfrac{{\mu^2 }}{{m^2 }}} \right) \right) \, \varepsilon
+ O(\varepsilon^2) \right] \,,\\ 
 B_0 \left( {m_1^2 ,m_2^2 } \right) = \dfrac{{A_0 \left( {m_1^2 } \right) - A_0 \left( {m_2^2 } \right)}}{{m_1^2  - m_2^2 }} \,,\\ 
 C_0 \left( {m_1^2 ,m_2^2 ,m_2^2 } \right) = \dfrac{{\partial B_0 \left( {m_1^2 ,m_2^2 } \right)}}{{\partial m_2^2 }} \,,\\ 
 B_0 \left( {p^2 ;m^2,0 } \right) = \dfrac{1}{\varepsilon } - \ln
 \left( {\dfrac{{m^2 }}{{\mu ^2 }}} \right) + 2 + \dfrac{{m^2  - p^2
 }}{{p^2 }}\ln \left( {\dfrac{{m^2  - p^2 - i0 }}{{m^2 }}} \right) +O(\varepsilon) \,,\\ 
 B_1 \left( {p^2 ;m^2,0 } \right) = 
\dfrac{1}{2 \, p^2} \, \left[A_0(m^2) - (p^2+m^2) \, B_0 \left(
  p^2;m^2,0 \right) \right] \,,\\ 
 B_1 \left( {p^2 ;0,m^2 } \right) = 
\dfrac{1}{2 \, p^2} \, \left[- A_0(m^2) - (p^2-m^2) \, B_0 \left(
  p^2;m^2,0 \right) \right] \,,\\ 
  B_1(0;m_1^2,m_2^2)   = - \dfrac{1}{2} \,
\left( {\dfrac{1}{\varepsilon }  + \ln \left( {x_\mu ^2 } \right) + \dfrac{{3 - 4x^2  + x^4  + \left( {4x^2  - 2x^4 } \right)\ln \left( {x^2 } \right)}}{{2\left( {1 - x^2 } \right)^2 }}} \right)\,,
 \end{array}
\end{equation}
with $x=m_2/m_1$.

\subsection{One-loop renormalization and counterterms}

\subsubsection{One-loop counterterm diagrams}
\label{sec:counter-term}

Squark-mass counterterm diagram:
\begin{equation}
\renewcommand{\arraystretch}{2}
\begin{array}{l}
 \Sigma _{q_f q_i }^{LR\;m\tilde{q}_{CT} }  = \dfrac{\alpha_s}{{2\pi }}C_F \, m_{\tilde g} \sum\limits_{s = 1}^6 {\delta m_{\tilde q_s }^2 } W_{fs}^{\tilde d} W_{i + 3s}^{\tilde d\star} C_0 \left( {m_{\tilde q_s }^2 ,m_{\tilde q_s }^2 ,m_{\tilde g}^2 } \right) \\ 
 \phantom{ \Sigma _{q_f q_i }^{LR\;\tilde g_{CT} } } = \dfrac{\alpha_s}{{2\pi }}C_F \, m_{\tilde g} \sum\limits_{s = 1}^6 {W_{fs}^{\tilde d} W_{i + 3s}^{\tilde d\star} \dfrac{{\delta m_{\tilde q_s }^2 }}{{m_{\tilde g}^2 }}\dfrac{1}{{\left( {1 - x_s^2 } \right)^2 }}\left[ {\ln \left( {x_s^2 } \right) + 1 - x_s^2 } \right.}  \\ 
\qquad \qquad + \left. {\varepsilon \left( {\ln \left( {x_\mu ^2 } \right)\left( {1 - x_s^2  + \ln \left( {x_s^2 } \right)} \right) - \dfrac{1}{2}\ln ^2 \left( {x_s^2 } \right) + \left( {\ln \left( {x_s^2 } \right) - 1} \right)x_s^2  + 1} \right)} \right] \\ 
 \end{array}
\end{equation}

Gluino mass counterterm diagram:
\begin{equation}
\renewcommand{\arraystretch}{2}
\begin{array}{l}
 \Sigma _{q_f q_i }^{LR\;m\tilde{g}_{CT} }  = \dfrac{\alpha_s}{{2\pi }}C_F \, \delta _{m_{\tilde g} } \sum\limits_{s = 1}^6 {W_{fs}^{\tilde d} W_{i + 3s}^{\tilde d\star} \left( {B_0 \left(m_{\tilde g}^2,  m_{\tilde q_s }^2  \right) + 2m_{\tilde g}^2 C_0 \left( {m_{\tilde q_s }^2 ,m_{\tilde g}^2 ,m_{\tilde g}^2 } \right)} \right)}  \\ 
\phantom{ \Sigma _{q_f q_i }^{LR\;\tilde q_{CT} }}  =
\dfrac{\alpha_s}{2\pi} C_F \, \delta _{m_{\tilde g} } \sum\limits_{s
  = 1}^6 W_{fs}^{\tilde d} W_{i + 3s}^{\tilde d\star} 
\left[ \dfrac{-x_s^2 \left( (1+x_s^2) \, \ln (x_s^2) + 2 \,(1-x_s^2) 
\right)}{(1-x_s^2)^2} \right. \\
\qquad \qquad -\varepsilon \dfrac{x_s^2}{2(1-x_s^2)^2} \left[
4 \, (1-x_s^2) +2 \, (1+x_s^2) \, \ln(x_s^2) 
-(1+x_s^2) \ln^2(1-x_s^2)
\right. \\
\left. \left.
\qquad \qquad \qquad \qquad \qquad
+\left( 4 \, (1-x_s^2) + 2 \, (1+x_s^2) \, \ln(x_s^2) \right)
\ln(x_{\mu}^2) \right] \right]
 \end{array}
\end{equation}
$\alpha_s$ counterterm diagram 
\begin{equation}
\renewcommand{\arraystretch}{2}
\begin{array}{l}
 \Sigma _{q_f q_i }^{LR\;\alpha_{s \;CT} }  = \dfrac{\delta
   \alpha_s}{{2\pi }}C_F \, m_{\tilde g} \sum\limits_{s = 1}^6
        {W_{fs}^{\tilde d} W_{i + 3s}^{\tilde d\star} B_0 \left( m_{\tilde g}^2, m_{\tilde q_s }^2   \right)}  \\ 
\phantom{ \Sigma _{q_f q_i }^{LR\;\alpha _{CT} } }  = \dfrac{{\delta
    \alpha _s }}{{2\pi }}C_F \, m_{\tilde{g}} \sum \limits_{s=1}^6 W_{fs}^{\tilde q} W_{i + 3,s}^{\tilde q\star} \left[ \dfrac{{x_s^2 \ln \left( {x_s^2 } \right)}}{{1 - x_s^2 }} \right.\\
\qquad\qquad\;\;\left.- \varepsilon \dfrac{{x_s^2 }}{{2\left( {1- x_s^2} \right)}}\left( {\ln ^2 \left( {x_s^2 } \right) - 2\ln \left( {x_s^2 } \right) - 2\ln \left( {x_s^2 } \right)\ln \left( {x_\mu ^2 } \right)} \right) \right] \\ 
 \end{array}
\end{equation}

\subsubsection{Renormalization of the Yukawa couplings in the MSSM}
Because of supersymmetry, the renormalization of the Yukawa coupling in
the quark-quark-Higgs vertex $Y^{q_i}$ 
and the one in squark-squark-Higgs vertex $Y^{\tilde{q}_i}$ must
be identical\footnote{This also includes that the renormalization of
  the Yukawa coupling entering the squark mass matrices is the same as
 the renormalization of the quark-quark-Higgs coupling.}. 
Indeed, we explicitly find that the counterterms for these couplings
are the same
\begin{equation}
Y_{}^{q_i,\tilde{q}_i\left( 0 \right)}  = Y_{}^{q_i,\tilde{q}_i}  +
\delta Y^{q_i,\tilde{q}_i} ,\;\;\;\delta Y^{q_i,\tilde{q}_i}  =
-\dfrac{{ \alpha _s}}{{4\pi }}\dfrac{2}{\varepsilon} C_F
Y^{q_i,\tilde{q}_i} \, ,
\label{Y_div}
\end{equation}
which even holds in the $\overline{\rm MS}$~scheme and in the $\overline{\rm DR}$~scheme at the one-loop level.
\subsubsection{$A$~term renormalization}

In the approximation $m_q=0$ the SQCD renormalization of the $A$-terms
is the same as of the Yukawa coupling\footnote{If $m_q\neq0$ the
  quark-gluino correction to $A$-terms induced flavor-non-diagonal
  (divergent) corrections.}.

\subsubsection{Squark mass renormalization}
We write the connection between the squares of bare and the renormalized
squark masses as 
\begin{equation}
\left(m^0_{\tilde{q}_t}\right)^2 =  
\left(m_{\tilde{q}_t}\right)^2 + 
\delta m_{\tilde{q}_t}^2 \, .
\end{equation}
From \eq{squark_gluon_SE}, \eq{squark_gluino_SE}, and
\eq{squark_squark_SE} and by taking into account that the second term
of \eq{squark_squark_SE} only renormalizes the Yukawa coupling (and the
$A$, $A'$ terms), 
we can easily read of $\delta m_{\tilde{q}_t}^2$. We obtain in the
$\overline{\rm MS}$~scheme:
\begin{equation}
\delta m_{\tilde{q_t}}^2 = \dfrac{\alpha_s}{4\pi} \, C_F \,
m_{\tilde{g}}^2 \, \left( (x_t^2+4) - x_t^2 \right) \, \dfrac{1}{\varepsilon} \, ,
\end{equation}
where the contribution proportional to $(x_t^2+4)$ comes from
\eq{squark_gluon_SE} and \eq{squark_gluino_SE} while the term $- x_t^2$ stems from the part of
\eq{squark_squark_SE} proportional to $\delta_{st}$.

\subsubsection{Gluino-mass renormalization}\label{sec:gluinorenormalization}
We decompose the gluino self-energy according to \eq{self-energy-decomposition}. Expressing the bare mass (marked with the superscript $(0)$) in terms of the physical one
\begin{equation}\begin{split}
& m_{\tilde{g}}^0 = m_{\tilde{g}} + \delta m_{\tilde{g}} \, ,
\end{split}\end{equation}
we get in the on-shell scheme
\begin{equation}
\delta m_{\tilde{g}} = -m_{\tilde{g}} \, \Sigma_{\tilde g}^{LL}(m_{\tilde{g}}^2)- \Sigma_{\tilde g}^{LR}(m_{\tilde{g}}^2) \, .
\label{gluinomassrenorm}
\end{equation}
For details see Ref.~\cite{Greub:2011ji}.
In the $\overline{\rm MS}$~scheme only the divergence of the
right-hand side enters: i.e., we get in this scheme
\begin{equation}
\delta m_{\tilde{g}} = - \dfrac{\alpha_s}{4\pi} \, m_{\tilde{g}} \, 
\left( 3 \, C_A - 2 \, tr \, n_f \right)  \, \dfrac{1}{\varepsilon} \, . 
\end{equation}

\subsubsection{Renormalization of $g_{s}$ in the MSSM}\label{gGren}
In lowest order, the strong coupling constant involved in $C^{q\,LR}_{fi}$ is
Yukawa type. The relation between the bare and the
renormalized version reads $g_{s,Y}^0=(1+\delta
Z_{g_{s,Y}})g_{s,Y}$, where the renormalization constant in the $\overline{\rm MS}$~scheme is given by
\begin{equation}
\begin{split}
\delta Z_{g_{s,Y}} =&\dfrac{\alpha_s}{4\pi} 
\left[ tr \, n_f - \dfrac{3}{2} C_A \right]
\dfrac{1}{\varepsilon} \, .
\end{split}
\end{equation}
Note that at one loop the renormalization constant is the same for the $\overline{\rm MS}$~scheme and the $\overline{\rm DR}$~scheme. 

\end{appendix}

\bibliography{2-loop-SE} 

\begin{thebibliography}{38}
\expandafter\ifx\csname natexlab\endcsname\relax\def\natexlab#1{#1}\fi
\expandafter\ifx\csname bibnamefont\endcsname\relax
  \def\bibnamefont#1{#1}\fi
\expandafter\ifx\csname bibfnamefont\endcsname\relax
  \def\bibfnamefont#1{#1}\fi
\expandafter\ifx\csname citenamefont\endcsname\relax
  \def\citenamefont#1{#1}\fi
\expandafter\ifx\csname url\endcsname\relax
  \def\url#1{\texttt{#1}}\fi
\expandafter\ifx\csname urlprefix\endcsname\relax\def\urlprefix{URL }\fi
\providecommand{\bibinfo}[2]{#2}
\providecommand{\eprint}[2][]{\url{#2}}

\bibitem[{\citenamefont{Hall et~al.}(1994)\citenamefont{Hall, Rattazzi, and
  Sarid}}]{Hall:1993gn}
\bibinfo{author}{\bibfnamefont{L.~J.} \bibnamefont{Hall}},
  \bibinfo{author}{\bibfnamefont{R.}~\bibnamefont{Rattazzi}}, \bibnamefont{and}
  \bibinfo{author}{\bibfnamefont{U.}~\bibnamefont{Sarid}},
  \bibinfo{journal}{Phys. Rev.} \textbf{\bibinfo{volume}{D50}},
  \bibinfo{pages}{7048} (\bibinfo{year}{1994}), \eprint{hep-ph/9306309}.

\bibitem[{\citenamefont{Carena et~al.}(1994)\citenamefont{Carena, Olechowski,
  Pokorski, and Wagner}}]{Carena:1994bv}
\bibinfo{author}{\bibfnamefont{M.~S.} \bibnamefont{Carena}},
  \bibinfo{author}{\bibfnamefont{M.}~\bibnamefont{Olechowski}},
  \bibinfo{author}{\bibfnamefont{S.}~\bibnamefont{Pokorski}}, \bibnamefont{and}
  \bibinfo{author}{\bibfnamefont{C.}~\bibnamefont{Wagner}},
  \bibinfo{journal}{Nucl.Phys.} \textbf{\bibinfo{volume}{B426}},
  \bibinfo{pages}{269} (\bibinfo{year}{1994}), \eprint{hep-ph/9402253}.

\bibitem[{\citenamefont{Carena et~al.}(2000)\citenamefont{Carena, Garcia,
  Nierste, and Wagner}}]{Carena:1999py}
\bibinfo{author}{\bibfnamefont{M.~S.} \bibnamefont{Carena}},
  \bibinfo{author}{\bibfnamefont{D.}~\bibnamefont{Garcia}},
  \bibinfo{author}{\bibfnamefont{U.}~\bibnamefont{Nierste}}, \bibnamefont{and}
  \bibinfo{author}{\bibfnamefont{C.~E.~M.} \bibnamefont{Wagner}},
  \bibinfo{journal}{Nucl. Phys.} \textbf{\bibinfo{volume}{B577}},
  \bibinfo{pages}{88} (\bibinfo{year}{2000}), \eprint{hep-ph/9912516}.

\bibitem[{\citenamefont{Bobeth et~al.}(2001)\citenamefont{Bobeth, Ewerth,
  Kruger, and Urban}}]{Bobeth:2001sq}
\bibinfo{author}{\bibfnamefont{C.}~\bibnamefont{Bobeth}},
  \bibinfo{author}{\bibfnamefont{T.}~\bibnamefont{Ewerth}},
  \bibinfo{author}{\bibfnamefont{F.}~\bibnamefont{Kruger}}, \bibnamefont{and}
  \bibinfo{author}{\bibfnamefont{J.}~\bibnamefont{Urban}},
  \bibinfo{journal}{Phys.Rev.} \textbf{\bibinfo{volume}{D64}},
  \bibinfo{pages}{074014} (\bibinfo{year}{2001}), \eprint{hep-ph/0104284}.

\bibitem[{\citenamefont{Isidori and Retico}(2001)}]{Isidori:2001fv}
\bibinfo{author}{\bibfnamefont{G.}~\bibnamefont{Isidori}} \bibnamefont{and}
  \bibinfo{author}{\bibfnamefont{A.}~\bibnamefont{Retico}},
  \bibinfo{journal}{JHEP} \textbf{\bibinfo{volume}{11}}, \bibinfo{pages}{001}
  (\bibinfo{year}{2001}), \eprint{hep-ph/0110121}.

\bibitem[{\citenamefont{Buras et~al.}(2003)\citenamefont{Buras, Chankowski,
  Rosiek, and Slawianowska}}]{Buras:2002vd}
\bibinfo{author}{\bibfnamefont{A.~J.} \bibnamefont{Buras}},
  \bibinfo{author}{\bibfnamefont{P.~H.} \bibnamefont{Chankowski}},
  \bibinfo{author}{\bibfnamefont{J.}~\bibnamefont{Rosiek}}, \bibnamefont{and}
  \bibinfo{author}{\bibfnamefont{L.}~\bibnamefont{Slawianowska}},
  \bibinfo{journal}{Nucl. Phys.} \textbf{\bibinfo{volume}{B659}},
  \bibinfo{pages}{3} (\bibinfo{year}{2003}), \eprint{hep-ph/0210145}.

\bibitem[{\citenamefont{Hofer et~al.}(2009)\citenamefont{Hofer, Nierste, and
  Scherer}}]{Hofer:2009xb}
\bibinfo{author}{\bibfnamefont{L.}~\bibnamefont{Hofer}},
  \bibinfo{author}{\bibfnamefont{U.}~\bibnamefont{Nierste}}, \bibnamefont{and}
  \bibinfo{author}{\bibfnamefont{D.}~\bibnamefont{Scherer}},
  \bibinfo{journal}{JHEP} \textbf{\bibinfo{volume}{10}}, \bibinfo{pages}{081}
  (\bibinfo{year}{2009}), \eprint{0907.5408}.

\bibitem[{\citenamefont{Babu and Kolda}(2000)}]{Babu:1999hn}
\bibinfo{author}{\bibfnamefont{K.}~\bibnamefont{Babu}} \bibnamefont{and}
  \bibinfo{author}{\bibfnamefont{C.~F.} \bibnamefont{Kolda}},
  \bibinfo{journal}{Phys.Rev.Lett.} \textbf{\bibinfo{volume}{84}},
  \bibinfo{pages}{228} (\bibinfo{year}{2000}), \eprint{hep-ph/9909476}.

\bibitem[{\citenamefont{Dedes and Pilaftsis}(2003)}]{Dedes:2002er}
\bibinfo{author}{\bibfnamefont{A.}~\bibnamefont{Dedes}} \bibnamefont{and}
  \bibinfo{author}{\bibfnamefont{A.}~\bibnamefont{Pilaftsis}},
  \bibinfo{journal}{Phys.Rev.} \textbf{\bibinfo{volume}{D67}},
  \bibinfo{pages}{015012} (\bibinfo{year}{2003}), \eprint{hep-ph/0209306}.

\bibitem[{\citenamefont{Crivellin and Nierste}(2009)}]{Crivellin:2008mq}
\bibinfo{author}{\bibfnamefont{A.}~\bibnamefont{Crivellin}} \bibnamefont{and}
  \bibinfo{author}{\bibfnamefont{U.}~\bibnamefont{Nierste}},
  \bibinfo{journal}{Phys. Rev.} \textbf{\bibinfo{volume}{D79}},
  \bibinfo{pages}{035018} (\bibinfo{year}{2009}), \eprint{0810.1613}.

\bibitem[{\citenamefont{Crivellin}(2010)}]{Crivellin:2009sd}
\bibinfo{author}{\bibfnamefont{A.}~\bibnamefont{Crivellin}},
  \bibinfo{journal}{Phys. Rev.} \textbf{\bibinfo{volume}{D81}},
  \bibinfo{pages}{031301} (\bibinfo{year}{2010}), \eprint{0907.2461}.

\bibitem[{\citenamefont{Diaz-Cruz et~al.}(2002)\citenamefont{Diaz-Cruz,
  Murayama, and Pierce}}]{DiazCruz:2000mn}
\bibinfo{author}{\bibfnamefont{J.}~\bibnamefont{Diaz-Cruz}},
  \bibinfo{author}{\bibfnamefont{H.}~\bibnamefont{Murayama}}, \bibnamefont{and}
  \bibinfo{author}{\bibfnamefont{A.}~\bibnamefont{Pierce}},
  \bibinfo{journal}{Phys.Rev.} \textbf{\bibinfo{volume}{D65}},
  \bibinfo{pages}{075011} (\bibinfo{year}{2002}), \eprint{hep-ph/0012275}.

\bibitem[{\citenamefont{Antusch and Spinrath}(2009)}]{Antusch:2009gu}
\bibinfo{author}{\bibfnamefont{S.}~\bibnamefont{Antusch}} \bibnamefont{and}
  \bibinfo{author}{\bibfnamefont{M.}~\bibnamefont{Spinrath}},
  \bibinfo{journal}{Phys.Rev.} \textbf{\bibinfo{volume}{D79}},
  \bibinfo{pages}{095004} (\bibinfo{year}{2009}), \eprint{0902.4644}.

\bibitem[{\citenamefont{Crivellin et~al.}(2011)\citenamefont{Crivellin, Hofer,
  and Rosiek}}]{Crivellin:2011jt}
\bibinfo{author}{\bibfnamefont{A.}~\bibnamefont{Crivellin}},
  \bibinfo{author}{\bibfnamefont{L.}~\bibnamefont{Hofer}}, \bibnamefont{and}
  \bibinfo{author}{\bibfnamefont{J.}~\bibnamefont{Rosiek}},
  \bibinfo{journal}{JHEP} \textbf{\bibinfo{volume}{1107}}, \bibinfo{pages}{017}
  (\bibinfo{year}{2011}), \eprint{1103.4272}.

\bibitem[{\citenamefont{Guasch et~al.}(2003)\citenamefont{Guasch, Hafliger, and
  Spira}}]{Guasch:2003cv}
\bibinfo{author}{\bibfnamefont{J.}~\bibnamefont{Guasch}},
  \bibinfo{author}{\bibfnamefont{P.}~\bibnamefont{Hafliger}}, \bibnamefont{and}
  \bibinfo{author}{\bibfnamefont{M.}~\bibnamefont{Spira}},
  \bibinfo{journal}{Phys. Rev.} \textbf{\bibinfo{volume}{D68}},
  \bibinfo{pages}{115001} (\bibinfo{year}{2003}), \eprint{hep-ph/0305101}.

\bibitem[{\citenamefont{Noth and Spira}(2011)}]{Noth:2010jy}
\bibinfo{author}{\bibfnamefont{D.}~\bibnamefont{Noth}} \bibnamefont{and}
  \bibinfo{author}{\bibfnamefont{M.}~\bibnamefont{Spira}},
  \bibinfo{journal}{JHEP} \textbf{\bibinfo{volume}{06}}, \bibinfo{pages}{084}
  (\bibinfo{year}{2011}), \eprint{1001.1935}.

\bibitem[{\citenamefont{Bauer et~al.}(2009)\citenamefont{Bauer, Mihaila, and
  Salomon}}]{Bauer:2008bj}
\bibinfo{author}{\bibfnamefont{A.}~\bibnamefont{Bauer}},
  \bibinfo{author}{\bibfnamefont{L.}~\bibnamefont{Mihaila}}, \bibnamefont{and}
  \bibinfo{author}{\bibfnamefont{J.}~\bibnamefont{Salomon}},
  \bibinfo{journal}{JHEP} \textbf{\bibinfo{volume}{02}}, \bibinfo{pages}{037}
  (\bibinfo{year}{2009}), \eprint{0810.5101}.

\bibitem[{\citenamefont{Bednyakov et~al.}(2003)\citenamefont{Bednyakov,
  Onishchenko, Velizhanin, and Veretin}}]{Bednyakov:2002sf}
\bibinfo{author}{\bibfnamefont{A.}~\bibnamefont{Bednyakov}},
  \bibinfo{author}{\bibfnamefont{A.}~\bibnamefont{Onishchenko}},
  \bibinfo{author}{\bibfnamefont{V.}~\bibnamefont{Velizhanin}},
  \bibnamefont{and} \bibinfo{author}{\bibfnamefont{O.}~\bibnamefont{Veretin}},
  \bibinfo{journal}{Eur. Phys. J.} \textbf{\bibinfo{volume}{C29}},
  \bibinfo{pages}{87} (\bibinfo{year}{2003}), \eprint{hep-ph/0210258}.

\bibitem[{\citenamefont{Bednyakov}(2010)}]{Bednyakov:2009wt}
\bibinfo{author}{\bibfnamefont{A.~V.} \bibnamefont{Bednyakov}},
  \bibinfo{journal}{Int. J. Mod. Phys.} \textbf{\bibinfo{volume}{A25}},
  \bibinfo{pages}{2437} (\bibinfo{year}{2010}), \eprint{0912.4652}.

\bibitem[{\citenamefont{Crivellin}(2011)}]{Crivellin:2010er}
\bibinfo{author}{\bibfnamefont{A.}~\bibnamefont{Crivellin}},
  \bibinfo{journal}{Phys. Rev.} \textbf{\bibinfo{volume}{D83}},
  \bibinfo{pages}{056001} (\bibinfo{year}{2011}), \eprint{1012.4840}.

\bibitem[{\citenamefont{Chatrchyan et~al.}(2011)}]{Collaboration:2011wc}
\bibinfo{author}{\bibfnamefont{S.}~\bibnamefont{Chatrchyan}}
  \bibnamefont{et~al.} (\bibinfo{collaboration}{CMS}) (\bibinfo{year}{2011}),
  \eprint{1103.0953}.

\bibitem[{\citenamefont{da~Costa et~al.}(2011)}]{daCosta:2011qk}
\bibinfo{author}{\bibfnamefont{J.~B.~G.} \bibnamefont{da~Costa}}
  \bibnamefont{et~al.} (\bibinfo{collaboration}{Atlas}) (\bibinfo{year}{2011}),
  \eprint{1102.5290}.

\bibitem[{\citenamefont{Gorbahn et~al.}(2011)\citenamefont{Gorbahn, Jager,
  Nierste, and Trine}}]{Gorbahn:2009pp}
\bibinfo{author}{\bibfnamefont{M.}~\bibnamefont{Gorbahn}},
  \bibinfo{author}{\bibfnamefont{S.}~\bibnamefont{Jager}},
  \bibinfo{author}{\bibfnamefont{U.}~\bibnamefont{Nierste}}, \bibnamefont{and}
  \bibinfo{author}{\bibfnamefont{S.}~\bibnamefont{Trine}},
  \bibinfo{journal}{Phys.Rev.} \textbf{\bibinfo{volume}{D84}},
  \bibinfo{pages}{034030} (\bibinfo{year}{2011}), \eprint{0901.2065}.

\bibitem[{\citenamefont{Blazek et~al.}(1995)\citenamefont{Blazek, Raby, and
  Pokorski}}]{Blazek:1995nv}
\bibinfo{author}{\bibfnamefont{T.}~\bibnamefont{Blazek}},
  \bibinfo{author}{\bibfnamefont{S.}~\bibnamefont{Raby}}, \bibnamefont{and}
  \bibinfo{author}{\bibfnamefont{S.}~\bibnamefont{Pokorski}},
  \bibinfo{journal}{Phys. Rev.} \textbf{\bibinfo{volume}{D52}},
  \bibinfo{pages}{4151} (\bibinfo{year}{1995}), \eprint{hep-ph/9504364}.

\bibitem[{\citenamefont{Crivellin and Greub}(2012)}]{Chargino}
\bibinfo{author}{\bibfnamefont{A.}~\bibnamefont{Crivellin}} \bibnamefont{and}
  \bibinfo{author}{\bibfnamefont{C.}~\bibnamefont{Greub}}, \bibinfo{journal}{in
  preparation}  (\bibinfo{year}{2012}).

\bibitem[{\citenamefont{Borzumati et~al.}(2000)\citenamefont{Borzumati, Greub,
  Hurth, and Wyler}}]{Borzumati:1999qt}
\bibinfo{author}{\bibfnamefont{F.}~\bibnamefont{Borzumati}},
  \bibinfo{author}{\bibfnamefont{C.}~\bibnamefont{Greub}},
  \bibinfo{author}{\bibfnamefont{T.}~\bibnamefont{Hurth}}, \bibnamefont{and}
  \bibinfo{author}{\bibfnamefont{D.}~\bibnamefont{Wyler}},
  \bibinfo{journal}{Phys.Rev.} \textbf{\bibinfo{volume}{D62}},
  \bibinfo{pages}{075005} (\bibinfo{year}{2000}), \eprint{hep-ph/9911245}.

\bibitem[{\citenamefont{Besmer et~al.}(2001)\citenamefont{Besmer, Greub, and
  Hurth}}]{Besmer:2001cj}
\bibinfo{author}{\bibfnamefont{T.}~\bibnamefont{Besmer}},
  \bibinfo{author}{\bibfnamefont{C.}~\bibnamefont{Greub}}, \bibnamefont{and}
  \bibinfo{author}{\bibfnamefont{T.}~\bibnamefont{Hurth}},
  \bibinfo{journal}{Nucl. Phys.} \textbf{\bibinfo{volume}{B609}},
  \bibinfo{pages}{359} (\bibinfo{year}{2001}), \eprint{hep-ph/0105292}.

\bibitem[{\citenamefont{Hall et~al.}(1986)\citenamefont{Hall, Kostelecky, and
  Raby}}]{Hall:1985dx}
\bibinfo{author}{\bibfnamefont{L.~J.} \bibnamefont{Hall}},
  \bibinfo{author}{\bibfnamefont{V.~A.} \bibnamefont{Kostelecky}},
  \bibnamefont{and} \bibinfo{author}{\bibfnamefont{S.}~\bibnamefont{Raby}},
  \bibinfo{journal}{Nucl.Phys.} \textbf{\bibinfo{volume}{B267}},
  \bibinfo{pages}{415} (\bibinfo{year}{1986}).

\bibitem[{\citenamefont{Crivellin and Girrbach}(2010)}]{Crivellin:2010gw}
\bibinfo{author}{\bibfnamefont{A.}~\bibnamefont{Crivellin}} \bibnamefont{and}
  \bibinfo{author}{\bibfnamefont{J.}~\bibnamefont{Girrbach}},
  \bibinfo{journal}{Phys. Rev.} \textbf{\bibinfo{volume}{D81}},
  \bibinfo{pages}{076001} (\bibinfo{year}{2010}), \eprint{1002.0227}.

\bibitem[{\citenamefont{Hahn}(2001)}]{Hahn:2000kx}
\bibinfo{author}{\bibfnamefont{T.}~\bibnamefont{Hahn}},
  \bibinfo{journal}{Comput. Phys. Commun.} \textbf{\bibinfo{volume}{140}},
  \bibinfo{pages}{418} (\bibinfo{year}{2001}), \eprint{hep-ph/0012260}.

\bibitem[{\citenamefont{Hahn and Schappacher}(2002)}]{Hahn:2001rv}
\bibinfo{author}{\bibfnamefont{T.}~\bibnamefont{Hahn}} \bibnamefont{and}
  \bibinfo{author}{\bibfnamefont{C.}~\bibnamefont{Schappacher}},
  \bibinfo{journal}{Comput. Phys. Commun.} \textbf{\bibinfo{volume}{143}},
  \bibinfo{pages}{54} (\bibinfo{year}{2002}), \eprint{hep-ph/0105349}.

\bibitem[{\citenamefont{Smirnov}(1995)}]{Smirnov:1994tg}
\bibinfo{author}{\bibfnamefont{V.~A.} \bibnamefont{Smirnov}},
  \bibinfo{journal}{Mod.Phys.Lett.} \textbf{\bibinfo{volume}{A10}},
  \bibinfo{pages}{1485} (\bibinfo{year}{1995}), \eprint{hep-th/9412063}.

\bibitem[{\citenamefont{Martin and Vaughn}(1993)}]{Martin:1993yx}
\bibinfo{author}{\bibfnamefont{S.~P.} \bibnamefont{Martin}} \bibnamefont{and}
  \bibinfo{author}{\bibfnamefont{M.~T.} \bibnamefont{Vaughn}},
  \bibinfo{journal}{Phys.Lett.} \textbf{\bibinfo{volume}{B318}},
  \bibinfo{pages}{331} (\bibinfo{year}{1993}), \eprint{hep-ph/9308222}.

\bibitem[{\citenamefont{Mihaila}(2009)}]{Mihaila:2009bn}
\bibinfo{author}{\bibfnamefont{L.}~\bibnamefont{Mihaila}},
  \bibinfo{journal}{Phys.Lett.} \textbf{\bibinfo{volume}{B681}},
  \bibinfo{pages}{52} (\bibinfo{year}{2009}), \eprint{0908.3403}.

\bibitem[{\citenamefont{Buras}(1998)}]{Buras:1998raa}
\bibinfo{author}{\bibfnamefont{A.~J.} \bibnamefont{Buras}}
  (\bibinfo{year}{1998}), \eprint{hep-ph/9806471}.

\bibitem[{\citenamefont{Carena et~al.}(2001)\citenamefont{Carena, Garcia,
  Nierste, and Wagner}}]{Carena:2000uj}
\bibinfo{author}{\bibfnamefont{M.~S.} \bibnamefont{Carena}},
  \bibinfo{author}{\bibfnamefont{D.}~\bibnamefont{Garcia}},
  \bibinfo{author}{\bibfnamefont{U.}~\bibnamefont{Nierste}}, \bibnamefont{and}
  \bibinfo{author}{\bibfnamefont{C.~E.~M.} \bibnamefont{Wagner}},
  \bibinfo{journal}{Phys. Lett.} \textbf{\bibinfo{volume}{B499}},
  \bibinfo{pages}{141} (\bibinfo{year}{2001}), \eprint{hep-ph/0010003}.

\bibitem[{\citenamefont{Degrassi et~al.}(2000)\citenamefont{Degrassi, Gambino,
  and Giudice}}]{Degrassi:2000qf}
\bibinfo{author}{\bibfnamefont{G.}~\bibnamefont{Degrassi}},
  \bibinfo{author}{\bibfnamefont{P.}~\bibnamefont{Gambino}}, \bibnamefont{and}
  \bibinfo{author}{\bibfnamefont{G.~F.} \bibnamefont{Giudice}},
  \bibinfo{journal}{JHEP} \textbf{\bibinfo{volume}{12}}, \bibinfo{pages}{009}
  (\bibinfo{year}{2000}), \eprint{hep-ph/0009337}.

\bibitem[{\citenamefont{Greub et~al.}(2011)\citenamefont{Greub, Hurth, Pilipp,
  Schupbach, and Steinhauser}}]{Greub:2011ji}
\bibinfo{author}{\bibfnamefont{C.}~\bibnamefont{Greub}},
  \bibinfo{author}{\bibfnamefont{T.}~\bibnamefont{Hurth}},
  \bibinfo{author}{\bibfnamefont{V.}~\bibnamefont{Pilipp}},
  \bibinfo{author}{\bibfnamefont{C.}~\bibnamefont{Schupbach}},
  \bibnamefont{and}
  \bibinfo{author}{\bibfnamefont{M.}~\bibnamefont{Steinhauser}},
  \bibinfo{journal}{Nucl. Phys.} \textbf{\bibinfo{volume}{B853}},
  \bibinfo{pages}{240} (\bibinfo{year}{2011}), \eprint{1105.1330}.

\end{thebibliography}

\end{document}